\documentclass[aps,pre,twocolumn,floatfix]{revtex4-2}

\usepackage{graphicx}
\usepackage{amsmath,empheq}
\usepackage{amssymb}
\usepackage{ifthen}
\usepackage{booktabs}

\usepackage{graphicx}
\usepackage{dcolumn}
\usepackage{bm}
\usepackage{gensymb}

\newcommand{\bb}{\begin{equation}}
\newcommand{\ee}{\end{equation}}
\newcommand{\ba}{\begin{eqnarray*}}
\newcommand{\ea}{\end{eqnarray*}}

\makeatletter
\def\@frameeq#1{%
  \framebox{$\,\displaystyle#1\hbox{\vrule height 2.4ex depth 1.4ex width 0pt}\,$}}
\newcommand\Equation[1]{$$\refstepcounter{equation}%
  \@frameeq{#1}%
  \eqno \hbox{\@eqnnum}$$\@ignoretrue\ignorespaces}
\newcommand\Displaystyle[1]{$$\@frameeq{#1}$$\@ignoretrue\ignorespaces}
\makeatother

\usepackage{longtable}
\usepackage[caption=false]{subfig}
\usepackage[allcolors=black,colorlinks]{hyperref}
 \usepackage{nameref}

\begin{document}

\title{Derjaguin-corrected Kelvin equation for finite open slits}

\author{Alexandr \surname{Malijevský}}
\affiliation{Research group of Molecular and Mesoscopic Modelling, The Czech Academy of Sciences, Institute of Chemical Process Fundamentals, 165 02 Prague, Czech Republic}
\affiliation{Department of Physical Chemistry, University of Chemistry and Technology Prague, 166 28 Prague, Czech Republic}


\begin{abstract}
We formulate a Derjaguin-corrected Kelvin equation for phase transitions in finite open slits. In such pores, capillary condensation is modified by the finite wall length because the condensed phase is bounded by two terminal menisci pinned at the slit ends. Starting from the macroscopic generalized Kelvin equation, we introduce a scaling approximation which replaces the implicit edge-contact-angle condition by a closed analytic expression while preserving the exact limiting behaviours. We then incorporate mesoscopic wetting-layer contributions into the finite-slit free-energy balance. The single-wall contribution naturally separates into a planar Derjaguin term and a residual finite-length correction controlled by the competition between the wall length $H$ and the parallel correlation length $\xi_\parallel$. The resulting equation reduces to the standard Derjaguin-Kelvin equation for infinitely long slits and to the macroscopic finite-slit Kelvin equation when adsorbed layers are absent. Explicit forms are derived for long-ranged dispersion forces and for short-ranged forces. The theory is tested using classical density functional theory for finite slits with attractive Lennard-Jones walls and hard walls, corresponding respectively to capillary condensation and its drying analogue, capillary evaporation. The Derjaguin-corrected theory substantially improves on the purely macroscopic prediction, capturing both the correct infinite-slit limit and the observed dependence on wall length. For the short-ranged hard-wall system, the agreement with DFT remains nearly quantitative even for very narrow slits.
\end{abstract}

\maketitle

\section{Introduction}

The properties of a fluid in the presence of an external field may differ markedly from those of its bulk counterpart \cite{Rowlinson, Henderson2009, Evans1990, Gelb2000, Binder2003}. The fundamental qualitative difference is that the one-body density is no longer uniform, but varies throughout the system according to the local density profile $\rho({\bf r})$. This non-uniformity gives rise to a wide range of interfacial and finite-size phenomena, which have been studied extensively over the last few decades, both theoretically and experimentally \cite{israel, Saam_2009, Bonn2009}.

One of the most striking consequences of confinement or external fields is the modification of phase behaviour. Even simple inhomogeneous systems may exhibit phase transitions with no direct bulk analogue. A basic example is a fluid in contact with a planar solid wall, which acts as a static external field, so that the equilibrium density varies in the direction perpendicular to the substrate, $\rho({\bf r})=\rho(z)$. The wall-fluid interaction then gives rise to wetting phenomena, associated with the formation of a liquid-like film at the wall-gas interface \cite{Cahn1977, Ebner_1977, Nakanishi1983, Dietrich1988, Schick,  Bonn2009}. Above the wetting temperature $T_w$, the wall is completely wet by the liquid phase and the equilibrium Young contact angle $\theta$ of a macroscopic droplet vanishes. In this complete-wetting regime, a thick wetting layer is adsorbed at the wall and its thickness $\ell_\pi$ increases as the pressure $p$ or chemical potential $\mu$ approaches its saturation value $p_{\rm sat}$ and $\mu_{\rm sat}$, respectively. In particular, the limit of $\delta p=p_{\rm sat}-p\to0$ (or $\delta\mu=\mu_{\rm sat}-\mu\to0$) is associated with the divergence of the film thickness $\ell_\pi\sim\delta p^{-\beta_{\rm s}^{\rm co}}$ and the parallel correlation length $\xi_\parallel\sim\delta p^{-\nu_\parallel^{\rm co}}$   \cite{Nakanishi1983, Dietrich_1985, Lipowsky_1985, Dietrich1991}. In three-dimensional systems these critical exponents are non-universal depending on the range of the forces: for long-ranged (LR) dispersion interactions $\beta_{\rm s}^{\rm co}=1/3$ and $\nu_\parallel^{\rm co}=2/3$, whereas for systems with only short-ranged (SR) interactions $\beta_{\rm s}^{\rm co}=0(\ln)$ and  $\nu_\parallel^{\rm co}=1/2$ \cite{Schick}.

Another elementary inhomogeneous system is a slit pore formed by two parallel walls separated by a distance $L$~\cite{Fisher1981,Evans1984,Evans_1986,Binder2008,Huber_2015,Bruschi2017}. In this geometry the density profile remains one-dimensional, but the combination of wall-fluid interactions and finite-size effects shifts the liquid-gas transition away from bulk coexistence. For walls which favour the liquid phase, this shift corresponds to capillary condensation, occurring at a pressure $p_{cc}(L)=p_{\rm sat}-\delta p_{cc}$, or equivalently at a chemical potential $\mu_{cc}(L)=\mu_{\rm sat}-\delta\mu_{cc}$. According to the classical Kelvin equation~\cite{Gregg}, the pressure shift is
\begin{equation}
	\delta p_{cc}^{\rm K}(L)=\frac{2\gamma\cos\theta}{L},
	\label{kelvin}
\end{equation}
where $\gamma$ is the liquid-gas surface tension. Thus, for hydrophilic walls, $\theta<\pi/2$, capillary condensation occurs before bulk condensation is reached. Equation~\eqref{kelvin} can be derived by balancing the free energies of the low-density gas-like and high-density liquid-like phases, but it also has a simple geometric interpretation: at coexistence the two phases may be separated by a circular meniscus of Laplace radius $R=\gamma/\delta p_{cc}$, which meets the walls at the Young contact angle $\theta$.

The Kelvin equation has proved to be quantitatively accurate even for surprisingly narrow slits, provided the walls are only partially wet \cite{Powles_1985, Mitropoulos_2008, Kim_2018, Yang2020}. However, in the complete-wetting regime its accuracy deteriorates because it does not account for the wetting films adsorbed at the walls. Their impact can be included by the Derjaguin correction \cite{Derjaguin1940, Evans_1985}, which replaces the geometric slit width by an effective width reduced by the adsorbed films,
\begin{equation}
	\delta p_{\rm cc}^{\rm DK}(L)=\frac{2\gamma\cos\theta}{L-n\ell_\pi}.
	\label{kelvin_derj}
\end{equation}
Here, $\ell_\pi$ is the planar wetting-layer thickness evaluated at the shifted coexistence point, so that Eq.~\eqref{kelvin_derj} must be solved self-consistently. The  factor $n$ depends on the range of the interactions with $n=3$ for long-ranged (LR) dispersion forces and $n=2$ for short-ranged (SR) forces.

A further extension of the simple geometrical models represent open capillaries formed of a pair of parallel walls of finite  length $H$ \cite{Malijevsky2017, Malijevsky2021a, Malijevsky2021}. Here, the density profile  depends on two coordinates, $\rho({\bf r})=\rho(x,z)$, while translational invariance is retained along the wall depth. The finite length has an immediate macroscopic consequence, such that the condensed phase must be separated from the surrounding reservoir by two terminal menisci. Compared with an infinite slit, these menisci add an interfacial free-energy cost and shift condensation closer to bulk saturation. Since the menisci are pinned at the open ends of the slit, they are not constrained to meet the walls at the Young contact angle. Instead, they are characterized by an edge contact angle $\theta_e$, which depends on the aspect ratio $L/H$.

In this paper, we formulate a Derjaguin-corrected Kelvin equation for capillary condensation in finite open slits. We first recall the macroscopic theory in terms of the edge contact angle of the terminal menisci and introduce a simple scaling approximation which converts the resulting implicit equation into a closed analytic form. This approximation preserves the exact limiting behaviours and is shown to be practically indistinguishable from the numerical solution of the full macroscopic equation. We then extend the theory to completely wet walls by incorporating the free-energy contribution of the adsorbed wetting layers. The resulting Derjaguin-corrected equation is the central result of the paper. It reduces to the standard Derjaguin-Kelvin equation in the limit of infinitely long slits, $H\to\infty$, and to the macroscopic finite-slit Kelvin equation when adsorbed layers are absent.  Combining this equation with the same scaling approximation yields a particularly simple form applicable to both LR and SR interactions. This is also relevant beyond the planar finite-slit geometry itself; the finite-slit construction provides the elementary building block for Kelvin-like descriptions of bridging transitions between non-planar walls, where the bridged region can be mapped onto an effective open slit \cite{Malijevsky2024, Malijevsky2024a}. The present Derjaguin-corrected formulation therefore offers a route to incorporate wetting- or drying-layer effects into such mapping approaches.

The theory is tested against microscopic density functional theory (DFT) calculations for finite slits formed by attractive Lennard-Jones walls and by purely hard walls. These two models represent, respectively, LR systems in the complete wetting regime and SR systems in the complete drying regime. In the former case the relevant transition is capillary condensation from the gas side, while in the latter it is capillary evaporation from the liquid side. The comparison shows that the Derjaguin correction substantially improves the prediction of the transition shift relative to the purely macroscopic finite-slit Kelvin equation, not only in the infinite-slit limit but also in the full dependence on the wall length. For the SR hard-wall system the agreement is particularly good, remaining very accurate even for narrow slits.

The rest of the paper is organized as follows. In Sec.~II we recall the macroscopic generalized Kelvin equation for finite slits and derive the scaling approximation, which provides a simple explicit alternative to the implicit edge contact angle equation. In Sec.~III we formulate the mesoscopic Derjaguin correction by analyzing the wetting layer on a single finite wall and then incorporating this contribution into the finite-slit free-energy balance. This leads to the Derjaguin-corrected Kelvin equation and to its scaling form for LR and SR forces. In Sec.~IV we define the microscopic DFT model and use it to test the theory. We first determine the planar wetting and drying film thicknesses and the infinite-slit limits, and then compare the finite-slit predictions with DFT results for capillary condensation in LR systems and capillary evaporation in SR systems. Finally, in Sec.~V we summarize the main results and discuss possible extensions of the theory.

\section{Macroscopic theory of capillary condensation in finite slits}

\subsection{Generalized Kelvin equation}

\begin{figure*}
	\centering
	\includegraphics[width=0.85\textwidth]{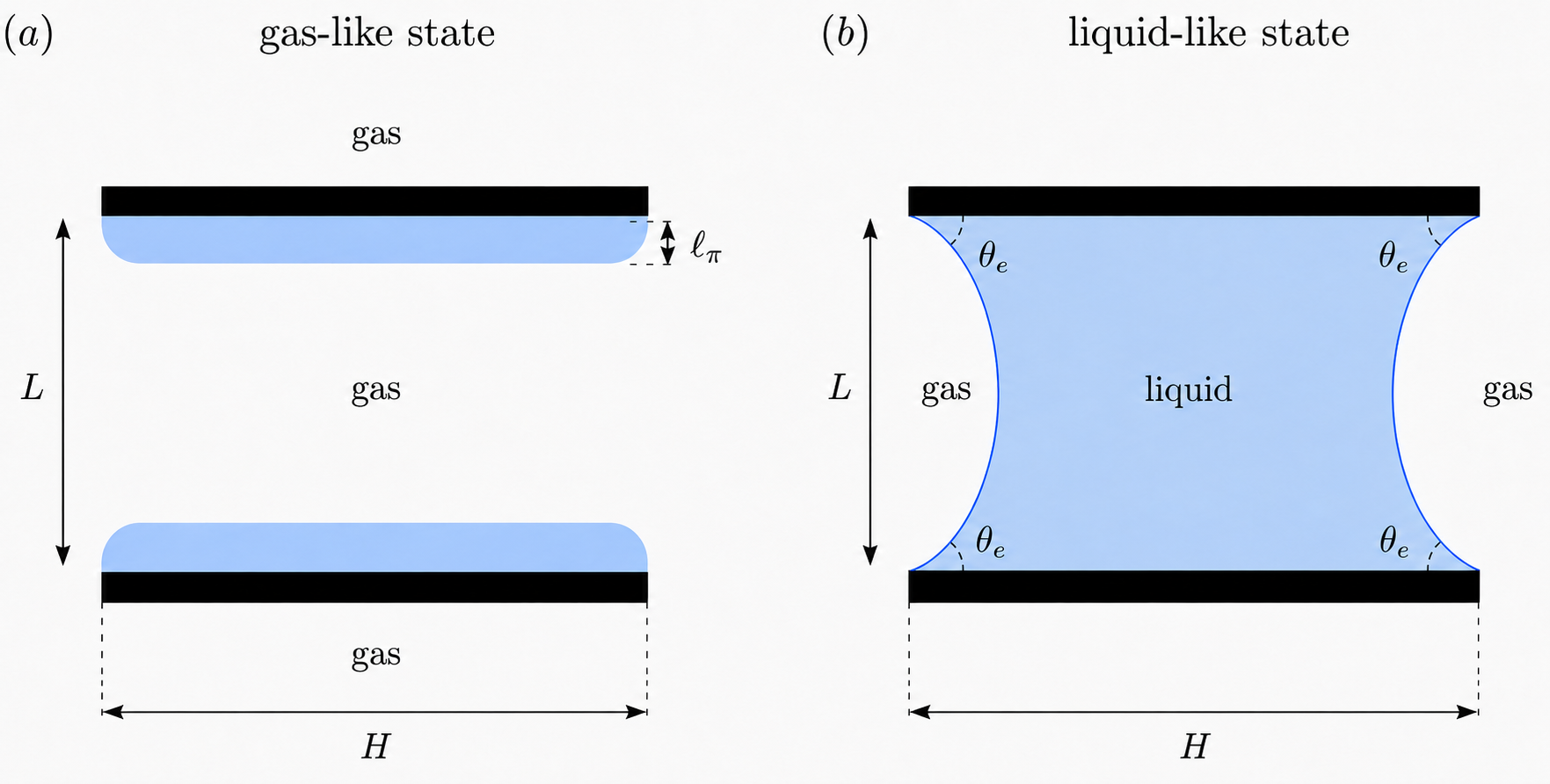}
	\caption{\label{fig:model_scheme}
		Schematic representation of the finite open slit and of the two competing states considered in the free-energy balance. 
		(a) Gas-like state for completely wet walls. The slit is filled by gas, while adsorbed liquid films of thickness $\ell_\pi$ coat the walls. These films are not included in the purely macroscopic theory of Sec.~II, but provide the mesoscopic contribution incorporated in Sec.~III.
		(b) Liquid-like state. The slit contains capillary liquid, separated from the surrounding gas reservoir by two terminal menisci pinned at the open ends of the walls. These menisci meet the walls at the edge contact angle $\theta_e$, which generally differs from the Young contact angle. The slit width is $L$ and the wall length is $H$.}
\end{figure*}

We begin with a brief review of the purely macroscopic arguments leading to the generalized Kelvin equation for capillary condensation in a finite open slit. 
We consider a slit of width $L$  formed by two identical parallel walls of length $H$ immersed in a bulk gas reservoir (see Fig.~\ref{fig:model_scheme}). The Kelvin equation for finite slits follows from a comparison of the grand potentials of the capillary gas-like and liquid-like states. For the gas-like phase, the grand potential per unit depth of the slit can be approximated as
\begin{equation}
	\Omega_g=-pLH+2\gamma_{wg}H,
\end{equation}
where $p$ is the pressure of the bulk gas reservoir and $\gamma_{wg}$ is the wall-gas surface tension. For the liquid-like phase, in addition to the pressure and wall-liquid surface contributions, one must include the free energy of the two terminal menisci of the Laplace radius $R$ connecting the walls at the edge contact angle $\theta_e$. Thus,
\begin{equation}
	\Omega_l=-p^\dagger(LH-2S)+2\gamma_{wl}H+2\gamma \ell ,
	\label{omega_l}
\end{equation}
where $p^\dagger=p-\delta p$ is the pressure of the metastable liquid and $\gamma_{wl}$ is the wall-liquid surface tension. Furthermore,  
\begin{equation}
	\ell=(\pi-2\theta_e)R,
\end{equation}
is the arc length of each meniscus, and
\begin{equation}
	S=\left(\frac{\pi}{2}-\theta_e\right)R^2-\frac{RL}{2}\sin\theta_e .
\end{equation}
 denotes the area between the meniscus and the open end of the slit \cite{Malijevsky2017}. 
The menisci are circular and geometry implies that their radius is
\begin{equation}
	R=\frac{L}{2\cos\theta_e}.
\end{equation}

Using Young's equation, one obtains the generalized Kelvin equation from the coexistence condition, $\Omega_g=\Omega_l$, which implies
\begin{equation} 
	\delta p_{cc}^{GK}(L;H)=\frac{2\gamma\cos\theta_e^{cc}}{L},
	\label{kelvinGK}
\end{equation}
where the edge contact angle at condensation, $\theta_e^{cc}$, satisfies the implicit relation \cite{Malijevsky2017}
\begin{equation}
	\cos\theta=\cos\theta_e^{cc}+\frac{L}{2H}
	\left[
	\sin\theta_e^{cc}+\left(\frac{\pi}{2}-\theta_e^{cc}\right)
	\sec\theta_e^{cc}
	\right].
	\label{theta_e_exact}
\end{equation}

This expression reduces continuously to the standard Kelvin equation in the macroscopic limit $H\to\infty$, for which $\theta_e^{cc}\to\theta^+$. As the
slit becomes shorter, $\theta_e^{cc}$ increases monotonically until the limiting case
 \bb
H^*=L\cos^{-1}\theta, \label{Lstar}
 \ee
when $\theta_e^{cc}=\pi/2$, corresponding to flat menisci and thus $\delta p_{cc}^{\rm GK}=0$. For shorter slits, $H<H^*$, the free-energy cost associated with the menisci exceeds the wall-fluid surface free-energy gain and capillary condensation is already suppressed.

\subsection{Scaling approximation}

Eq.~(\ref{kelvinGK}) provides the exact macroscopic condition for capillary condensation in a finite slit; however, it requires determination of the edge contact angle, given implicitly by Eq.~(\ref{theta_e_exact}) that must be solved numerically. Here, we introduce a simple approximation which converts the generalized Kelvin equation into an explicit closed form.

The exact theory contains two limiting constraints. Firstly, in the limit of very long slits, formally $L/H\to0$, the finite-slit correction must vanish and the standard Kelvin equation must be recovered,
\begin{equation}
	\delta p_{cc}^{GK}(L;H)\to\delta p_{\rm cc}^{\rm K}(L)=\frac{2\gamma\cos\theta}{L}.
\end{equation}

Secondly, capillary condensation is suppressed when the menisci become flat, corresponding to
\begin{equation}
	\theta_e^{cc}=\frac{\pi}{2},\qquad\delta p_{\rm cc}^{\rm GK}=0,
\end{equation}
which occurs when $L=L^*\equiv H\cos\theta$.

We now assume that the finite-size correction enters through the single scaling variable $x=L/L^*$ and write
\begin{equation}
	\delta p_{cc}^{\rm GK}(L;H)=\delta p_{\rm cc}^{\rm K}(L)\,W(x),
\end{equation}
where $W(x)$ is a dimensionless scaling function.

The two exact limits impose
\begin{equation}
	W(0)=1,\qquad W(1)=0.
\end{equation}
Moreover, the exact solution of the generalized Kelvin equation shows that the pressure shift vanishes linearly as $L\to L^{*-}$. Therefore the simplest admissible scaling form is
\begin{equation}
	W(x)=1-x,
\end{equation}
which yields the explicit approximation
\begin{equation}
		\delta p_{\rm cc}^{\rm GK}(L;H)\approx
		\frac{2\gamma\cos\theta}{L}-\frac{2\gamma}{H},
		\label{kelvin_scale}
\end{equation}
separating $L$ and $H$ dependence and corresponding, via Eq.~(\ref{kelvinGK}), to the approximation for the edge contact angle
\begin{equation}
	\theta_e\approx\arccos\left(\cos\theta-\frac{L}{H}\right).
	\label{theta_e_scale}
\end{equation}

\begin{figure}[t]
	\centering
	\includegraphics[width=\linewidth]{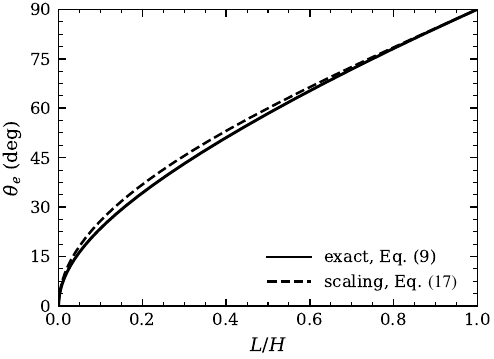}
\caption{\label{fig:thetae_scaling}
	Comparison between the exact edge contact angle $\theta_e$ obtained from the generalized Kelvin equation, Eq.~\eqref{theta_e_exact}, and its scaling approximation, Eq.~\eqref{theta_e_scale}, for complete wetting ($\theta=0$). By construction, the approximation reproduces the exact limits at $L/H\to0$ and $L/H\to1$, together with their leading asymptotic approaches. The resulting interpolation remains in close agreement with the exact solution over the entire range of aspect ratios.}

\end{figure}

By construction, the approximation is exact in both limiting cases: it reduces to the Kelvin equation for $L/H\to0$, and it gives $\delta p_{\rm cc}^{\rm GK}=0$ at $L=L^*$. Moreover, for complete wetting, $\theta=0$, Eq.~(\ref{theta_e_scale}) yields in the limit $L/H\ll1$
\begin{equation}
	\theta_e\sim\left(\frac{L}{H}\right)^{1/2},
\end{equation}
thus reproducing the correct square-root asymptotic dependence of $\theta_e$ implied by Eq.~(\ref{theta_e_exact}) \cite{Malijevsky2017}. At the opposite endpoint, $L/H\to1$, expansion of Eq.~(\ref{theta_e_scale}) gives
\begin{equation}
	\frac{\pi}{2}-\theta_e \sim 1-\frac{L}{H},
\end{equation}
which is again the correct leading asymptotic behaviour. Hence the scaling approximation reproduces not only the exact limiting values at both ends of the interval $0\le L/H\le1$, but also their leading approaches. It is therefore reasonable to expect that a smooth matching of these asymptotic regimes provides an accurate representation of the full solution. This expectation is confirmed in Fig.~\ref{fig:thetae_scaling}, which shows excellent agreement between the approximate and exact edge contact angles over the entire range of aspect ratios. Consequently, Eq.~(\ref{kelvin_scale}) provides a simple closed analytic approximation to the generalized Kelvin equation, avoiding the need to numerically solve the transcendental equation for $\theta_e^{cc}$.

\section{Mesoscopic extension: Derjaguin-like correction}

For $T>T_w$, the walls are completely wet and the gas-like state in the finite slit is no longer dry. Instead, each wall is coated by a mesoscopic wetting layer whose thickness depends on the distance from bulk saturation. The aim of this section is to formulate the corresponding mesoscopic correction to the generalized Kelvin equation.

We proceed in two steps. First, we consider the wetting layer adsorbed at a single finite wall of length $H$. This allows us to identify the residual finite-size contributions to the grand potential and their dependence on the length scales $H$, $\ell_\pi$, and $\xi_\parallel$. The resulting single-wall free energy will then be used in the next subsection to correct the finite-slit coexistence condition.

\subsection{Single finite wall}

\subsubsection{General formulation}

We consider a finite wall segment of length $H$ in contact with a bulk gas reservoir. In the complete wetting regime the wall is coated by a liquid-like film whose local thickness is denoted by $\ell(x)$, where $x$ is the coordinate along the wall. The mesoscopic contribution of this film is described by the interfacial Hamiltonian
\begin{equation}
	{\cal H}[\ell]=
	\int_0^H dx\,
	\left[
	\frac{\gamma}{2}(\ell')^2+W(\ell)
	\right],
	\label{eq:IH_single_wall}
\end{equation}
where $W(\ell)$ is the binding potential whose form depends on the range of the microscopic forces and will be specified below. This potential contains both the effective interaction between the wall and the liquid-gas interface of the wetting film and the free-energy cost of maintaining a metastable liquid film, proportional to $\delta p$.


For an infinite planar wall, the equilibrium film thickness $\ell_\pi$ at mean-field level is determined by
\begin{equation}
	W'(\ell_\pi)=0,
	\label{eq:ellpi_def}
\end{equation}
and the corresponding planar film free energy per unit length is
\begin{equation}
	f_\pi(\delta p)=\gamma_{wl}+\gamma+W(\ell_\pi).
	\label{eq:fpi_def}
\end{equation}
The associated parallel correlation length is
\begin{equation}
	\xi_\parallel
	=
	\left[
	\frac{\gamma}{W''(\ell_\pi)}
	\right]^{1/2},
	\label{eq:xipar_definition}
\end{equation}
which characterizes correlations along the planar wetting film. Equivalently, within the mean-field interfacial Hamiltonian, it is the lateral length over which a small perturbation of the film thickness relaxes back to its mean value $\ell_\pi$.

\begin{figure*}[t]
	\centering
	\includegraphics[width=\textwidth]{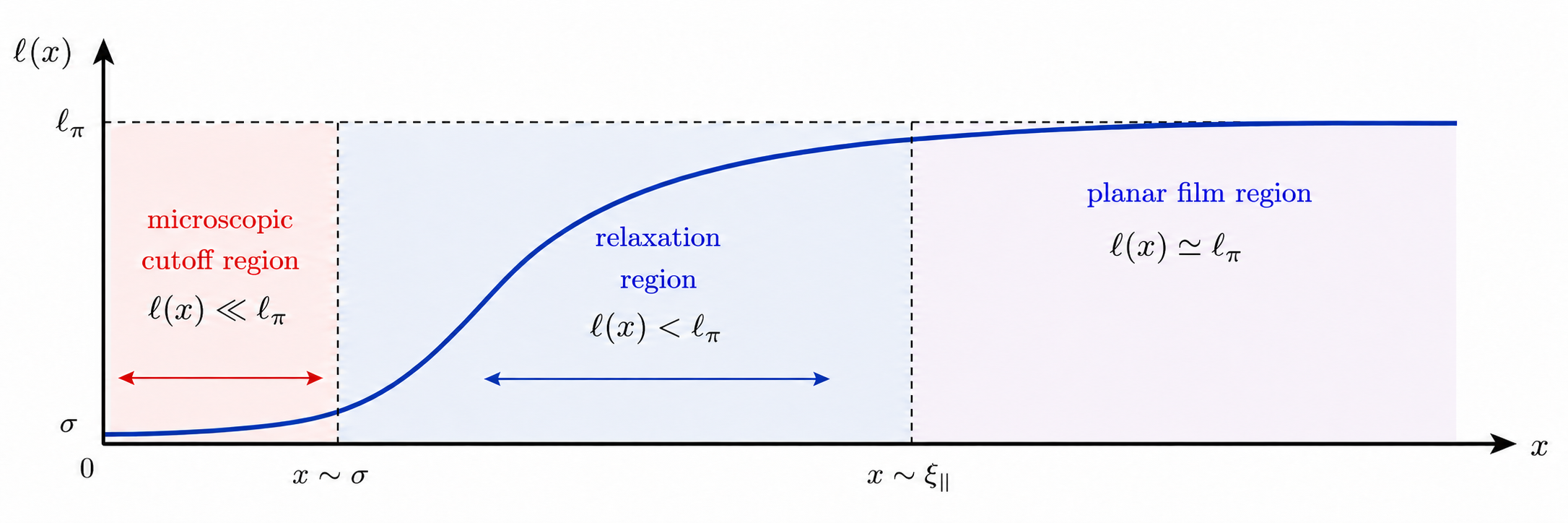}
\caption{\label{fig:edge_profile}
	Schematic structure of a wetting film near the edge of a finite completely wet wall. The microscopic cutoff region has width of order $\sigma$, where the interfacial Hamiltonian description is not expected to apply. The relaxation region connects this microscopic edge profile to the planar film region over the scale $\xi_\parallel$, during which $\ell(x)$ approaches the planar wetting-layer thickness $\ell_\pi$. On a sufficiently long wall, the remaining part of the film is essentially planar, $\ell(x)\simeq \ell_\pi$.}
\end{figure*}

The problem therefore involves three relevant length scales: the planar film thickness $\ell_\pi$, the parallel correlation length $\xi_\parallel$, and a microscopic cutoff $\sigma$, of the order of the molecular diameter, which sets the lower length scale below which the interfacial Hamiltonian description is no longer expected to apply.

For a finite wall, it is natural to separate the grand potential into the planar contribution and a residual finite-size part:
\begin{equation}
	\Omega_{\rm w}(H,\delta p)=
	Hf_\pi(\delta p)+\Delta\Omega_{\rm w}(H,\delta p).
	\label{eq:Omega_w_decomp_general}
\end{equation}

As indicated schematically in Fig.~\ref{fig:edge_profile}, the finite-wall profile contains edge regions where the film thickness changes from microscopic values to the planar thickness $\ell_\pi$. These edge regions may either be independent or strongly overlapping, depending on the ratio $H/\xi_\parallel$. We write
\begin{equation}
	\Delta\Omega_{\rm w}(H,\delta p)
	=
	2\tau_{\rm e}(\delta p)+{\cal F}_{\rm C}(H,\delta p),
	\label{eq:delta_om}
\end{equation}
 where the two non-extensive contributions have clear interpretations in the following limiting regimes.

\begin{itemize}
	
	\item {\bf Film regime} ($H\gg\xi_\parallel$). In this limit, the film reaches its planar thickness $\ell_\pi$ over a distance of order $\xi_\parallel$ from each edge. The two edge regions are asymptotically independent, so that ${\cal F}_{\rm C}\to0$ and the residual contribution reduces to
	\begin{equation}
		\Delta\Omega_{\rm w}(H,\delta p)\simeq
		2\tau_{\rm e}(\delta p),\qquad H\gg\xi_\parallel.
		\label{eq:R_film_regime}
	\end{equation}
	The edge free energy, or line-tension contribution, is then defined by the semi-infinite-wall limit
	\begin{equation}
		\tau_{\rm e}(\delta p)=
		\frac{1}{2} \lim_{H\to\infty}
		\left[\Omega_{\rm w}(H,\delta p)-Hf_\pi(\delta p)\right].
		\label{eq:tau_definition}
	\end{equation}

	\item {\bf Droplet regime} ($H\ll\xi_\parallel$). In the opposite case, the wall is too short for the profile to relax to $\ell_\pi$. The two edge regions overlap and form a single droplet-like object extending over the whole wall section. The residual contribution can no longer be interpreted as two independent edge terms. The deviation from this independent-edge limit defines the effective finite-size interaction
	\begin{equation}
		{\cal F}_{\rm C}(H,\delta p)=
		\Delta\Omega_{\rm w}(H,\delta p)-2\tau_{\rm e}(\delta p).
		\label{eq:Casimir_definition}
	\end{equation}
	We refer to ${\cal F}_{\rm C}$ as a Casimir-like interaction free energy, since it represents the effective interaction between the two edge regions induced by the finite wall length, with a range controlled by $\xi_\parallel$.

\end{itemize}

The explicit forms of $\tau_{\rm e}$ and ${\cal F}_{\rm C}$ depend on the range of the microscopic forces, which we specify next.

\subsubsection{Long-ranged dispersion forces}

For long-ranged dispersion forces the binding potential has the asymptotic form
\begin{equation}
	W_{\rm LR}(\ell)=\delta p\,\ell+\frac{A}{\ell^2},
	\label{eq:W_LR_single}
\end{equation}
where $A>0$ is the Hamaker constant. 
The equilibrium planar film thickness is obtained from $W_{\rm LR}'(\ell_\pi)=0$, giving
\begin{equation}
	\ell_\pi=\left(\frac{2A}{\delta p}\right)^{1/3},
	\label{eq:ell_LR_single}
\end{equation}
thus, at equilibrium,
\begin{equation}
	W_{\rm LR}(\ell_\pi)=\frac{3A}{\ell_\pi^2}=\frac{3}{2}\delta p\,\ell_\pi,
	\label{eq:Wpi_LR_single}
\end{equation}
and
\begin{equation}
	\xi_\parallel^{\rm LR}=\left(\frac{\gamma}{6A}\right)^{1/2}\ell_\pi^2\sim\delta p^{-2/3}.
	\label{eq:xi_LR_single}
\end{equation}

The mean-field equilibrium local height profile satisfies the Euler-Lagrange equation  
\begin{equation}
	\gamma \ddot{\ell}=W'(\ell),
	\label{eq:EL_single_wall}
\end{equation}
the exact solution of which is not available in a simple closed form. We therefore use the two asymptotic regimes to obtain the limiting forms of the residual free energy and then interpolate between them.

Consider first the film regime, $H\gg\xi_\parallel$, where the two edges are independent. The excess free energy of one edge is obtained by subtracting the planar film contribution:
\begin{equation}
	\tau_{\rm e}^{\rm LR}=
	\int dx\,\left[\frac{\gamma}{2}\dot{\ell}^2+
	W_{\rm LR}(\ell)-W_{\rm LR}(\ell_\pi)\right].
	\label{eq:tau_LR_exact_start}
\end{equation}
The first integral of the Euler-Lagrange equation is
\begin{equation}
	\frac{\gamma}{2}\dot{\ell}^2
	=W_{\rm LR}(\ell)-W_{\rm LR}(\ell_\pi),
	\label{eq:first_integral_full_LR}
\end{equation}
consistent with the condition $\ell\to\ell_\pi$ far from the edge. 

Using Eq.~(\ref{eq:first_integral_full_LR}), the edge free energy may be written as an integral over the film height,
\begin{equation}
	\tau_{\rm e}^{\rm LR}=
	\int_{\sigma}^{\ell_\pi} d\ell\,
	\left[2\gamma\left(W_{\rm LR}(\ell)-W_{\rm LR}(\ell_\pi)\right)\right]^{1/2}+{\cal{O}}(1),
	\label{eq:tau_height_integral_LR}
\end{equation}
where $\sigma$ is a microscopic cutoff and the omitted term is a contribution from the microscopic cutoff region.

Substituting $\delta p=2A/\ell_\pi^3$, which follows from \eqref{eq:ell_LR_single},  gives
\begin{equation}
	W_{\rm LR}(\ell)-W_{\rm LR}(\ell_\pi)
	=\frac{A}{\ell_\pi^2}\left[2u+\frac{1}{u^2}-3\right],
	\qquad u=\frac{\ell}{\ell_\pi},
	\label{eq:W_full_scaled_LR}
\end{equation}
which upon inserting into Eq.~(\ref{eq:tau_height_integral_LR}) gives
\begin{equation}
	\tau_{\rm e}^{\rm LR}=
	\sqrt{2A\gamma}\int_{\sigma/\ell_\pi}^{1}du\,
	\left[2u+\frac{1}{u^2}-3\right]^{1/2}
	+{\cal{O}}(1).
	\label{eq:tau_scaled_integral_LR}
\end{equation}
The dominant contribution to the integral comes from the lower limit,
\begin{equation}
	\left[2u+\frac{1}{u^2}-3\right]^{1/2}
	=\frac{1}{u}+{\cal{O}}(u),
\end{equation}
and therefore
\begin{equation}
	\tau_{\rm e}^{\rm LR}\approx
	\sqrt{2A\gamma}\,\ln\left(\frac{\ell_\pi}{\sigma}\right).
	\label{eq:tau_log_ellpi}
\end{equation}
Since $\xi_\parallel^{\rm LR}\propto \ell_\pi^2$, this may equivalently be written  as
\begin{equation}
	\tau_{\rm e}^{\rm LR}
	\simeq\frac{\alpha}{2}
	\ln\left(\frac{\xi_\parallel}{\sigma}\right),\qquad\alpha=\sqrt{2A\gamma}.
	\label{eq:tau_LR_final}
\end{equation}

The same result may also be expressed as an integral over the lateral coordinate. To see this, note that in the dominant part of the integral, $u=\ell/\ell_\pi\ll1$, Eq.~(\ref{eq:first_integral_full_LR}) reduces to
\begin{equation}
	\frac{\gamma}{2}\dot{\ell}^2
	\simeq
	\frac{A}{\ell^2}.
\end{equation}
Hence
\begin{equation}
	\ell\,d\ell
	=
	\left(\frac{2A}{\gamma}\right)^{1/2}dx,
\end{equation}
so that, to leading order,
\begin{equation}
	x\propto \ell^2,
	\qquad\frac{d\ell}{\ell}
	=\frac{1}{2}\frac{dx}{x}
\end{equation}
and thus
\begin{equation}
	\tau_{\rm e}^{\rm LR}
	\simeq\frac{\alpha}{2}
	\int_{\sigma}^{\xi_\parallel}\frac{dx}{x},
\end{equation}
reproducing \eqref{eq:tau_LR_final}.

Thus, in the film regime $H\gg\xi_\parallel$, the residual grand potential is
\begin{equation}
	\Delta\Omega_{\rm w}(H,\delta p)
	\simeq2\tau_{\rm e}^{\rm LR}
	\simeq\alpha\int_{\sigma}^{\xi_\parallel}\frac{dx}{x}
	=\alpha\ln\left(\frac{\xi_\parallel}{\sigma}\right),
	\qquad H\gg\xi_\parallel .
	\label{eq:R_LR_film}
\end{equation}

In the droplet regime,  the two edge regions overlap before either of them can relax to the planar film thickness $\ell_\pi$. The logarithmic contribution is therefore cut off by $H$ rather than by $\xi_\parallel$, i.e., to leading order,
\begin{equation}
	\Delta\Omega_{{\rm w}}^{\rm LR}(H)
	\simeq\alpha\int_{\sigma}^{H}\frac{dx}{x}
	=\alpha\ln\left(\frac{H}{\sigma}\right),\qquad H\ll\xi_\parallel .
	\label{eq:DeltaOmega_LR_drop_0}
\end{equation}

It remains to check that the pressure term gives only a subleading correction in this regime. Using the algebraic saturation profile, $\ell_0(x)\sim (A/\gamma)^{1/4}x^{1/2}$, its contribution scales as
\begin{equation}
	\Delta\Omega_p
	\sim\delta p\int_0^H dx\,\ell_0(x)
	\sim\delta p\left(\frac{A}{\gamma}\right)^{1/4}H^{3/2}
	\sim\alpha\left(\frac{H}{\xi_\parallel}\right)^{3/2},
	\label{eq:pressure_correction_drop}
\end{equation}
which for $H\ll\xi_\parallel$ indeed  does not change the leading logarithmic result. The droplet-regime residual grand potential is therefore
\begin{equation}
	\Delta\Omega_{\rm w}^{\rm LR}(H,\delta p)
	\simeq\alpha\ln\left(\frac{H}{\sigma}\right),
     \qquad H\ll\xi_\parallel .
	\label{eq:DeltaOmega_LR_drop}
\end{equation}

Using the definition
\begin{equation}
	{\cal F}_{\rm C}=\Delta\Omega_{\rm w}-2\tau_{\rm e},
\end{equation}
and the film-regime result $2\tau_{\rm e}^{\rm LR}\simeq\alpha\ln(\xi_\parallel/\sigma)$, we obtain
\begin{equation}
	{\cal F}_{\rm C}^{\rm LR}(H,\delta p)
	\simeq-\alpha\ln\left(\frac{\xi_\parallel}{H}\right).
	\label{eq:Casimir_LR_drop}
\end{equation}
Thus the Casimir-like interaction is attractive, lowering the residual grand potential relative to two independent edge regions.

Finally, the two limiting results can be connected by introducing an effective wall length
  \begin{equation}
	H_{\rm eff}=\frac{H}{\sqrt{1+(H/\xi_\parallel)^2}},
	\label{eq:Heff_single_wall}
\end{equation}
and writing
\begin{equation}
	\Delta\Omega_{\rm w}^{\rm LR}(H,\delta p)
	\simeq\alpha\ln\left(\frac{H_{\rm eff}}{\sigma}\right),
	\label{eq:R_LR_Heff}
\end{equation}
which smoothly interpolates between the two asymptotic results \eqref{eq:R_LR_film} and \eqref{eq:DeltaOmega_LR_drop_0}. 

Combining the planar extensive and residual terms, the total single-wall grand potential for long-ranged forces is therefore
\begin{equation}
	\Omega_{\rm w}^{\rm LR}(H,\delta p)\simeq
	H\left[\gamma_{wl}+\gamma+\frac{3A}{\ell_\pi^2}\right]
	+\alpha\ln\left(\frac{H_{\rm eff}}{\sigma}\right).
	\label{eq:Omega_w_LR_final}
\end{equation}

\subsubsection{Short-ranged forces}

If only short-ranged interactions are present, the binding potential has the asymptotic form
\begin{equation}
	W_{\rm SR}(\ell)
	=\delta p\ell+A e^{-\kappa \ell},
	\label{eq:W_SR}
\end{equation}
where $\kappa^{-1}=\xi_b$ is the bulk correlation length. 

For an infinite wall, minimization of (\ref{eq:W_SR}) yields
\begin{equation}
	\kappa\ell_\pi
	=\ln\left(\frac{\kappa A}{\delta p}\right).
	\label{eq:ell_SR}
\end{equation}
and thus  
\begin{equation}
	W_{\rm SR}(\ell_\pi)\simeq\delta p\ell_\pi.
	\label{eq:Wpi_SR}
\end{equation}
Close to saturation and away from $T_c$, both  $\ell_\pi$ and $\xi_\parallel$ are large compared with $\xi_b$ and Eq.~\eqref{eq:xipar_definition} gives
\begin{equation}
	\xi_\parallel^{\rm SR}
	=\left(\frac{\gamma}{\kappa\delta p}\right)^{1/2}.
	\label{eq:xi_SR}
\end{equation}

In the intermediate part of the edge profile, where $\ell$ is still well below $\ell_\pi$, the pressure term is subdominant. Indeed, using $\delta p=\kappa A e^{-\kappa\ell_\pi}$,
\begin{equation}
	\frac{\delta p\,\ell}{Ae^{-\kappa\ell}}
	=\kappa\ell\,e^{-\kappa(\ell_\pi-\ell)}\ll 1,
\end{equation}
provided $\ell_\pi-\ell$ exceeds a microscopic distance. Thus the local structure of the SR tail is governed by the saturation-like Hamiltonian
\begin{equation}
	{\cal H}_{\rm tail}^{\rm SR}
	=\int dx\left[\frac{\gamma}{2}\dot{\ell}^{\,2}+Ae^{-\kappa\ell}\right].
\end{equation}
Its first integral is
\begin{equation}
	\frac{\gamma}{2}\dot{\ell}^{\,2}=Ae^{-\kappa\ell},
\end{equation}

which implies
\begin{equation}
	\ell(x)\sim \frac{2}{\kappa}\ln\left(\frac{x}{\sigma}\right).
\end{equation}
Thus the SR tail grows only logarithmically with the distance from the edge. For $\delta p>0$, this saturation-like growth is terminated when $\ell(x)$ approaches $\ell_\pi$; the pressure term then becomes comparable to the exponential binding term and makes the profile relax to the planar film thickness. The corresponding lateral cutoff is of order $\xi_\parallel^{\rm SR}$; however, unlike for LR forces, this cutoff does not generate a logarithmic edge free energy. Using the first integral, the edge contribution scales as
\begin{equation}
	\tau_{\rm e}^{\rm SR}
	\sim(2\gamma A)^{1/2}\int_\sigma^{\ell_\pi}d\ell\,e^{-\kappa\ell/2}.
\end{equation}
This integral converges as $\ell_\pi\to\infty$ and the SR edge contribution thus remains finite even at saturation 
\begin{equation}
	\tau_{\rm e}^{\rm SR}={\cal{O}}(1)
\end{equation}
and represents a microscopic, nonuniversal finite-size contribution. In the finite-slit free-energy balance it appears only divided by $H$, giving a correction of order $\sigma/H$, and is beyond the leading Derjaguin order retained here.
Thus, to leading  order,
\begin{equation}
	\Delta\Omega_{\rm w}^{\rm SR}(H,\delta p)={\cal{O}}(1),
	\label{eq:DeltaOmega_SR_O1}
\end{equation}
and the dominant mesoscopic correction is the extensive planar contribution proportional to $H W_{\rm SR}(\ell_\pi)$. The single-wall grand potential for SR forces is therefore
\begin{equation}
	\Omega_{\rm w}^{\rm SR}(H,\delta p)
	\simeq H\left[\gamma_{wl}+\gamma+\delta p\,\ell_\pi\right],
	\label{eq:Omega_w_SR_final}
\end{equation}
where the omitted terms are subleading finite-size corrections.

We note that this decomposition is closely related to the
finite-size analysis of the line tension at wetting by Indekeu
and Dobbs~\cite{IndekeuDobbs1994}; see also Ref.~\cite{Indekeu1994}
for a broader review of line-tension singularities at wetting. In that problem the interfacial profile connects a microscopic thin-film region to
a mesoscopic thick wetting film, and the line tension is obtained
from the free energy of this transition zone. The boundary
conditions are therefore of the same type as those appearing
near an open edge in the present finite-wall geometry, where
the film thickness evolves from a microscopic value at the edge
towards the planar value $\ell_\pi$ over a distance of order
$\xi_\parallel$. In the present notation, $\tau_e$ is the
free-energy contribution of one such edge region, while
$F_C(H,\delta p)$ accounts for the finite-size interaction
between the two edge regions when their relaxation zones
overlap. This connection also explains why the long-ranged
case gives a logarithmic contribution cut off by $\xi_\parallel$
or by $H$, whereas for short-ranged forces the corresponding
edge term remains finite and microscopic.

\subsection{Derjaguin-corrected Kelvin equation for finite slits}

We now use the previous results to correct the free-energy balance in finite slits for $T>T_w$. The gas-like branch contains two finite wet walls and is described by
\begin{equation}
	\Omega_g^{\rm meso}=-pLH+2\Omega_w(H,\delta p),
\end{equation}
where
\begin{equation}
	\Omega_w(H,\delta p)=
	H(\gamma_{wl}+\gamma+W_\pi)+
	\Delta\Omega_w(H,\delta p),
\end{equation}
with $W_\pi\equiv W(\ell_\pi)$,
separating the planar Derjaguin contribution from the residual finite-wall contribution.

 For the liquid-like state we retain the macroscopic geometrical
construction, but supplement it by the line-tension contribution
associated with the four contact regions where the terminal menisci
meet the open wall edges. This puts the liquid-like branch on the same
footing as the gas-like branch, whose finite-wall free energy already
contains the edge contributions. Using the natural LR line-tension scale
$\alpha=\sqrt{2A\gamma}$, we estimate one contribution $\alpha$ per
corner and define  $\Omega_l^{\rm meso}=\Omega_l+4\alpha$.
For SR forces, the analogous line-tension contribution is microscopic, of order ${\cal{O}}(\gamma\sigma)$, and is neglected at the present level.


The coexistence condition $\Omega_g^{\rm meso}=\Omega_l^{\rm meso}$ then gives
\begin{equation}
	\delta p(LH-2S)=2\gamma H+2HW_\pi+2\Delta\Omega_{\rm w}-2\gamma\ell-4\alpha .
	\label{eq:DGK_balance}
\end{equation}
Substituting for the geometric measures given in Sec.~II~A this yields
\begin{align}
	1
	&=\cos\theta_e
	+\frac{L}{2H}\left[\sin\theta_e+\left(\frac{\pi}{2}-\theta_e
	\right)\sec\theta_e\right]\nonumber\\
	&\quad-\frac{W_\pi(\delta p)}{\gamma}
	-\frac{\Delta\Omega_{\rm w}(H,\delta p)}{\gamma H}+\frac{2\alpha}{\gamma H}.
	\label{eq:DGK_general}
\end{align}
Eq.~(\ref{eq:DGK_general}), together with
\begin{equation}
	\delta p_{cc}^{\rm DGK}=\frac{2\gamma\cos\theta_e}{L},
	\label{eq:DGK_pressure}
\end{equation}
is the general Derjaguin-corrected Kelvin equation for finite slits.

\subsubsection{Long-ranged dispersion forces}

For LR dispersion forces, $W_\pi^{\rm LR}=3A/\ell_\pi^2$ and $\Delta\Omega_{\rm w}^{\rm LR}(H,\delta p)\simeq\alpha\ln(H_{\rm eff}/\sigma)$, where $\alpha=\sqrt{2A\gamma}$. Using Eq.~(\ref{eq:DGK_pressure}) and $\ell_\pi=(2A/\delta p)^{1/3}$ gives
\begin{align}
	1
	&=\cos\theta_e+\frac{L}{2H}\left[
	\sin\theta_e+\left(\frac{\pi}{2}-\theta_e\right)
	\sec\theta_e\right]\nonumber\\
	&\quad-3\left(\frac{A\cos^2\theta_e}{\gamma L^2}\right)^{\frac{1}{3}}    
	-\frac{\alpha}{\gamma H}\ln\left(\frac{H_{\rm eff}}{\sigma}\right)+\frac{2\alpha}{\gamma H}.
	\label{eq:DGK_LR}
\end{align}

Here, the explicit pressure-dependence has been eliminated except for $H_{\rm eff}=H/\sqrt{1+(H/\xi_\parallel^{\rm LR})^2}$ as a compact notation, evaluated through
\begin{equation}
	 	\xi_\parallel^{\rm LR}=\left(\frac{\gamma}{6A}\right)^{1/2}\ell_\pi^2,\qquad  
	\ell_\pi=\left(\frac{AL}{\gamma\cos\theta_e}\right)^{1/3}.
\end{equation}

The infinite-slit limit is recovered immediately. For $H\to\infty$, both finite-height terms vanish and Eq.~(\ref{eq:DGK_LR}) reduces to
\begin{equation}
	1=\frac{\delta p L}{2\gamma}-\frac{3A}{\gamma\ell_\pi^2}.
	\label{eq:LR_limit_1}
\end{equation}
Using $3A/\ell_\pi^2=(3/2)\delta p\,\ell_\pi$, one obtains
\begin{equation}
	\delta p_{cc}^{\rm DK}=\frac{2\gamma}{L-3\ell_\pi},
	\label{eq:DK_LR_limit}
\end{equation}
which is the standard Derjaguin-Kelvin equation for an infinite slit with LR forces in the complete-wetting regime.

\subsubsection{Short-ranged forces}

For SR forces, $W_\pi^{\rm SR}\simeq\delta p\,\ell_\pi$, with $\ell_\pi=\kappa^{-1}\ln(\kappa A/\delta p)$, while $\Delta\Omega_{\rm w}^{\rm SR}={\cal O}(\gamma\sigma)$. Substitution into Eq.~(\ref{eq:DGK_general}) gives
\begin{equation}
	1=\cos\theta_e+\frac{L}{2H}\left[\sin\theta_e+
	\left(\frac{\pi}{2}-\theta_e\right)\sec\theta_e\right]
	-\frac{\delta p\,\ell_\pi}{\gamma}+{\cal{O}}\left(\frac{\sigma}{H}\right).
	\label{eq:DGK_SR_general}
\end{equation}
The ${\cal O}(\sigma/H)$ term is microscopic compared with the retained geometric finite-height correction ${\cal{O}}(L/H)$ and is neglected at the present Derjaguin level. Using Eq.~(\ref{eq:DGK_pressure}) gives the closed implicit equation for $\theta_e$:
\begin{align}
	1
	&=\cos\theta_e+\frac{L}{2H}\left[
	\sin\theta_e+\left(\frac{\pi}{2}-\theta_e\right)\sec\theta_e\right]\nonumber\\
	&-\frac{2\cos\theta_e}{L\kappa}\ln\left(\frac{L\kappa A}{2\gamma\cos\theta_e}\right).
	\label{eq:DGK_SR2}
\end{align}

For $H\to\infty$, Eq.~(\ref{eq:DGK_SR_general}) reduces to
\begin{equation}
	1=\frac{\delta p L}{2\gamma}-\frac{\delta p\,\ell_\pi}{\gamma},
	\label{eq:SR_limit_1}
\end{equation}
and hence
\begin{equation}
	\delta p_{cc}^{\rm DK}=\frac{2\gamma}{L-2\ell_\pi},
	\label{eq:DK_SR_limit}
\end{equation}
which is the standard Derjaguin-Kelvin equation for an infinite slit with SR forces in the complete-wetting regime.

\subsection{Scaling approximation to the mesoscopic Kelvin equation}

It is useful to combine the Derjaguin-corrected equation derived above with the scaling approximation of Sec.~II B. In the corresponding macroscopic approach, this approximation replaces, for $\theta=0$, the exact geometric contribution in Eq.~(\ref{eq:DGK_general}) by its linear scaling form,
\begin{equation}
	\frac{L}{2H}\left[\sin\theta_e+\left(\frac{\pi}{2}-\theta_e\right)\sec\theta_e\right]\longrightarrow \frac{L}{H}.
\end{equation}
Indeed, for $W_\pi=\Delta\Omega_{\rm w}=0$, this gives $\delta p\simeq 2\gamma L^{-1}(1-L/H)$, i.e. the scaling result of Sec.~IIB for $\theta=0$.

Applying the same approximation to Eq.~(\ref{eq:DGK_general}) gives
\begin{equation}
	\frac{\delta p L}{2\gamma}\simeq 1-\frac{L}{H}+\frac{W_\pi(\delta p)}{\gamma}+\frac{\Delta\Omega_{\rm w}(H,\delta p)}{\gamma H}-\frac{2\alpha}{\gamma H},
	\label{eq:DGK_scaling_general}
\end{equation}
with $W_\pi=3A/\ell_\pi^2$ and $\Delta\Omega_{\rm w}=\alpha\ln(H_{\rm eff}/\sigma)$ for LR forces and $W_\pi=\delta p\ell_\pi$ and $\Delta\Omega_{\rm w}=O(\gamma\sigma)$ for SR forces.

Eq.~(\ref{eq:DGK_scaling_general}) represents the scaling extension of the generalized Kelvin equation to completely wet finite slits and is a central result of this work. For sufficiently large slits, the wetting-layer thickness remains small compared with the slit width, $\ell_\pi\ll L$, and Eq.~\eqref{eq:DGK_scaling_general} can be recast in the following form
\begin{equation}
	\delta p\,[L-n\ell_\pi]
	\simeq
	2\gamma\left(1-\frac{L}{H}\right)+\frac{2\Delta\Omega_w}{H}-\frac{4\alpha}{H},
\end{equation}
which may be viewed as the finite-slit analogue of Eq.~\eqref{kelvin_derj}.

Specifically, for LR dispersion forces, $\Delta\Omega_{\rm w}^{\rm LR}\simeq\alpha\ln(H_{\rm eff}/\sigma)$, where $\alpha=\sqrt{2A\gamma}$. Thus
\begin{equation}
	\delta p\left[L-3\ell_\pi(\delta p)\right]\simeq 2\gamma\left(1-\frac{L}{H}\right)+\frac{2\alpha}{H}\ln\left(\frac{H_{\rm eff}}{\sigma}\right) -\frac{4\alpha}{H},
	\label{eq:DGK_scaling_LR}
\end{equation}
with $\ell_\pi=(2A/\delta p)^{1/3}$ and $H_{\rm eff}=H/[1+(H/\xi_\parallel^{\rm LR})^2]^{1/2}$.

For SR forces, $\Delta\Omega_{\rm w}^{\rm SR}=O(\gamma\sigma)$ and is neglected at the present level. Hence
\begin{equation}
	\delta p\left[L-2\ell_\pi(\delta p)\right]\simeq 2\gamma\left(1-\frac{L}{H}\right),
	\qquad
	\ell_\pi=\frac{1}{\kappa}\ln\left(\frac{\kappa A}{\delta p}\right).
	\label{eq:DGK_scaling_SR}
\end{equation}

Eqs.~(\ref{eq:DGK_scaling_LR}) and (\ref{eq:DGK_scaling_SR}) reduce to the standard Derjaguin-Kelvin equations in the limit $H\to\infty$. Conversely, setting $\ell_\pi=0$ and $\Delta\Omega_{\rm w}=0$ recovers the original scaling approximation given by Eq.~\eqref{kelvin_scale}. These compact forms recast the planar wetting contribution as an effective-width correction and should therefore be used away from the endpoint $L/H=1$. As $L/H\to1$, however, $\delta p_{cc}\to0$ and $\ell_\pi$ diverges, so that the more general scaling equation~(\ref{eq:DGK_scaling_general}) should be used directly.

\section{DFT test of the generalized Derjaguin-Kelvin equation}

In this section, we test the generalized Derjaguin-Kelvin equation against results obtained from classical density functional theory. We first define the microscopic model by specifying the fluid-fluid and wall-fluid interactions, and then outline the corresponding DFT formulation.  We subsequently determine the location of capillary condensation and evaporation in finite open slits with different widths and wall lengths and compare the numerical results, for both LR and SR systems, with the theoretical prediction based on the scaling form of Eq.~\eqref{eq:DGK_scaling_general}.

\subsection{Microscopic model and density functional theory}

We consider a one-component simple fluid confined by a finite open slit. The fluid-fluid interaction is represented by a hard core of diameter $\sigma$, supplemented by an attractive Lennard-Jones tail truncated at a cutoff distance $r_c$. The hard-core repulsion is incorporated through a hard-sphere reference system, while the attractive part is treated perturbatively, as specified below. The confining slit consists of two identical parallel finite wall segments whose inner surfaces are separated by a distance $L$ and have length $H$ in the $x$ direction. The lower wall surface lies at $z=0$ for $0<x<H$, while the upper wall surface lies at $z=L$ over the same interval. The solid domains themselves are semi-infinite in the $z$ direction: the lower wall occupies $0<x<H$, $z<0$, whereas the upper wall occupies $0<x<H$, $z>L$. Both walls are macroscopically extended in the $y$ direction, so that the equilibrium density profile is translationally invariant along $y$ and depends only on $x$ and $z$.

The equilibrium density profile is obtained by minimizing the grand-potential functional \cite{Evans1979}
\begin{equation}
	\Omega[\rho]=F[\rho]+\int d{\bf r}\,\rho({\bf r})[V({\bf r})-\mu],
	\label{eq:DFT_grand_potential}
\end{equation}
where $\mu$ is the chemical potential and $V({\bf r})$ is the external potential exerted by the confining walls. The intrinsic Helmholtz free-energy functional is written in the standard perturbative form
\begin{equation}
	F[\rho]=F_{\rm id}[\rho]+F_{\rm ex}[\rho]. 
	\label{eq:DFT_F_split}
\end{equation}
The  ideal-gas contribution is known exactly
\begin{equation}
	\beta F_{\rm id}[\rho]=\int d{\bf r}\,\rho({\bf r})\left\{\ln[\rho({\bf r})\Lambda^3]-1\right\},
	\label{eq:DFT_ideal}
\end{equation}
where $\Lambda$ is the thermal de Broglie wavelength, while the excess part of the free-energy functional is further separated 
\begin{equation}
 F_{\rm ex}[\rho]=F_{\rm hs}[\rho]+F_{\rm att}[\rho],
\end{equation} 
into a purely repulsive, hard-sphere contribution $F_{\rm hs}[\rho]$ and an attractive part $F_{\rm att}[\rho]$.

The hard-sphere part $F_{\rm hs}$ is approximated using Rosenfeld's fundamental-measure theory~\cite{Rosenfeld1989}. In this approach the excess free energy of the hard-sphere reference fluid is written as a local functional of a set of weighted densities,
\begin{equation}
	\beta F_{\rm hs}[\rho]=\int d{\bf r}\,\Phi(\{n_\alpha({\bf r})\}),
\end{equation}
where
\begin{equation}
	n_\alpha({\bf r})=\int d{\bf r}'\,\rho({\bf r}')w_\alpha({\bf r}-{\bf r}').
\end{equation}
For a one-component hard-sphere fluid of radius $R=\sigma/2$, the Rosenfeld weights are
\begin{align}
	w_3({\bf r})&=\Theta(R-r),&
	w_2({\bf r})&=\delta(R-r), \nonumber\\
	w_1({\bf r})&=\frac{w_2({\bf r})}{4\pi R},&
	w_0({\bf r})&=\frac{w_2({\bf r})}{4\pi R^2}, \nonumber\\
	{\bf w}_{v2}({\bf r})&=\frac{{\bf r}}{r}\delta(R-r),&
	{\bf w}_{v1}({\bf r})&=\frac{{\bf w}_{v2}({\bf r})}{4\pi R},
\end{align}
while the corresponding free-energy density is
\begin{equation}
	\Phi=
	-n_0\ln(1-n_3)
	+\frac{n_1n_2-{\bf n}_{v1}\cdot{\bf n}_{v2}}{1-n_3}
	+\frac{n_2^3-3n_2{\bf n}_{v2}\cdot{\bf n}_{v2}}{24\pi(1-n_3)^2}.
\end{equation}
This form accounts accurately for the packing effects induced by the hard core and reduces to the Percus-Yevick compressibility equation of state in the homogeneous limit. 

The attractive part of the fluid-fluid interaction is treated at mean-field level,
\begin{equation}
	F_{\rm att}[\rho]=\frac{1}{2}\int d{\bf r}\int d{\bf r}'\,\rho({\bf r})\rho({\bf r}')u_{\rm a}(|{\bf r}-{\bf r}'|),
	\label{eq:DFT_att}
\end{equation}
where
\begin{equation}
	u_{\rm a}(r)=
	\begin{cases}
		0, & r<\sigma,\\
		-4\varepsilon(\sigma/r)^6, & \sigma<r<r_c,\\
		0, & r>r_c,
	\end{cases}
	\label{eq:DFT_uatt}
\end{equation}
with $r_c=2.5\sigma$. The truncation of the attractive tail makes the fluid-fluid interaction effectively short ranged. In the following, $\varepsilon$ and $\sigma$ are used as the units of energy and length, respectively.

We use two different models for the wall-fluid interaction. In the first, LR model, the slit is embedded in a bulk gas. Each wall is formed by Lennard-Jones atoms distributed uniformly within the wall volume with number density $\rho_w$. A wall atom interacts with a fluid particle through
\begin{equation}
	\phi_w(r)=4\varepsilon_w\left[\left(\frac{\sigma}{r}\right)^{12}-\left(\frac{\sigma}{r}\right)^6\right].
	\label{eq:wall_atom_LJ}
\end{equation}
The walls also contain a hard core and are therefore impenetrable: for the lower wall the external potential is infinite for $0<x<H$ and $z<0$, while for the upper wall it is infinite for $0<x<H$ and $z>L$.

Outside the hard cores, the potential of the lower wall is obtained by integrating $\phi_w(r)$ over the wall volume $0<x'<H$, $-\infty<y'<\infty$, $z'<0$. It can be written as
\begin{equation}
	V_{\rm LJ}(x,z)=V_6(x,z)+V_{12}(x,z).
	\label{eq:wall_potential_split}
\end{equation}

The attractive contribution is
\begin{equation}
	\begin{split}
		V_6(x,z)=\alpha_w\bigg[&
		\psi_6(x-H,z)-\psi_6(x,z) \\
		&+\frac{1}{(x-H)^3}-\frac{1}{x^3}
		\bigg],
		\label{eq:wall_V6}
	\end{split}
\end{equation}
where $\alpha_w=-\pi\varepsilon_w\rho_w\sigma^6/3$. The repulsive contribution is
\begin{equation}
	\begin{split}
		V_{12}(x,z)=\pi\varepsilon_w\rho_w\sigma^{12}\bigg[&
		\psi_{12}(x-H,z)-\psi_{12}(x,z) \\
		&+\frac{2}{45}\left(\frac{1}{(x-H)^9}-\frac{1}{x^9}\right)
		\bigg].
		\label{eq:wall_V12}
	\end{split}
\end{equation}

The auxiliary functions are
\begin{equation}
	\psi_6(x,z)=-\frac{2x^4+x^2z^2+2z^4}{2z^3x^3(x^2+z^2)^{1/2}},
	\label{eq:psi6}
\end{equation}
and
  \begin{widetext}
\begin{equation}
	\psi_{12}(x,z)=-\frac{1}{2880}\frac{
		128x^{16}+448x^{14}z^2+560x^{12}z^4+280x^{10}z^6+35x^8z^8+280x^6z^{10}+560x^4z^{12}+448x^2z^{14}+128z^{16}
	}{
		z^9x^9(x^2+z^2)^{7/2}
	}.
	\label{eq:psi12}
\end{equation}
  \end{widetext}
The full external potential of the slit is obtained by adding the contribution from the upper wall by reflection about $z=L$,
\begin{equation}
	V_{\rm LR}(x,z)=V_{\rm LJ}(x,z)+V_{\rm LJ}(x,L-z),
	\label{eq:wall_full_LR}
\end{equation}
outside the hard cores. This model contains a long-ranged wall-fluid attraction and will be used to study capillary condensation in finite slits.

The second wall model is purely repulsive and is represented by hard walls. They are embedded in a bulk liquid, and therefore the relevant first-order transition is capillary evaporation rather than capillary condensation. The lower hard wall is represented by
\begin{equation}
	V_{\rm HW}(x,z)=
	\begin{cases}
		\infty, & 0<x<H,\ z<0,\\
		0, & \text{otherwise},
	\end{cases}
	\label{eq:hard_wall_bottom}
\end{equation}
and the upper hard wall is obtained by reflection about $z=L$. In this case, no LR interactions are present, so that this model represents the SR case with the full slit potential
\begin{equation}
	V_{\rm SR}(x,z)=V_{\rm HW}(x,z)+V_{\rm HW}(x,L-z).
	\label{eq:wall_full_SR}
\end{equation}

 \begin{figure}[t]
	\centering
	\includegraphics[width=\linewidth]{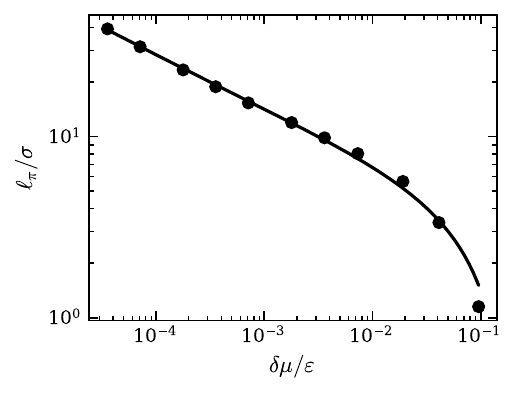}
	\includegraphics[width=\linewidth]{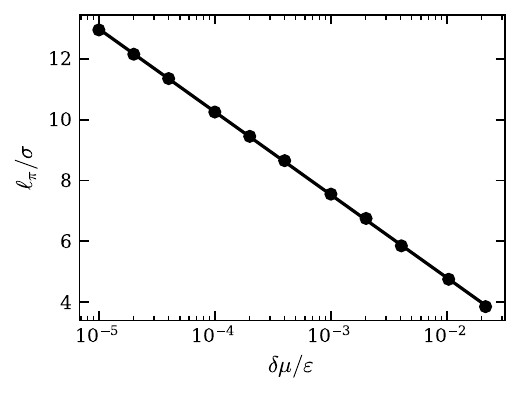}
	\caption{\label{fig:ell_pi}
		Planar wetting/drying layer thickness $\ell_\pi$ as a function of the departure from bulk coexistence, $\delta\mu=|\mu-\mu_{\rm sat}|$, obtained from one-dimensional DFT calculations at $T=0.92T_c$. (a) Complete wetting at an attractive Lennard-Jones $9$-$3$ wall, approached from the gas side, $\mu\to\mu_{\rm sat}^{-}$. The symbols denote DFT results and the solid line is a fit including the leading LR complete-wetting behaviour, $\ell_\pi\sim\delta\mu^{-1/3}$, together with subdominant corrections. (b) Complete drying at a hard wall, approached from the liquid side, $\mu\to\mu_{\rm sat}^{+}$. The semi-log plot confirms the logarithmic SR asymptotic form $\ell_\pi\simeq-\xi_b\ln\delta\mu$, with $\xi_b=1.184\sigma$.}
\end{figure}

\begin{figure}[t]
	\centering
	\includegraphics[width=\linewidth]{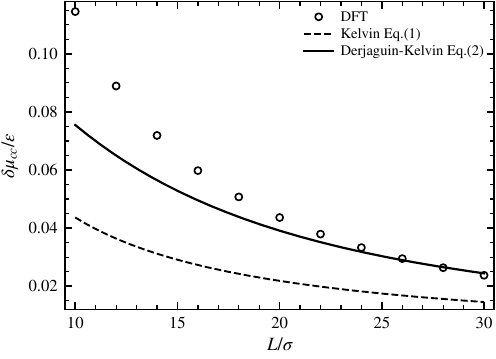}
	\caption{\label{fig:CC_inf_slit}
		Capillary condensation shift $\delta\mu_{cc}$ in an infinite slit formed by two infinite Lennard-Jones walls at $T=0.92T_c$. Symbols denote DFT results. The dashed line is the prediction of the classical Kelvin equation, Eq.~(\ref{kelvin}), while the solid line is the Derjaguin-Kelvin equation, Eq.~(\ref{kelvin_derj}), with $n=3$ appropriate for LR dispersion forces. The Derjaguin correction significantly improves the agreement with DFT over the entire range of slit widths $L$.}
\end{figure}

\begin{figure}[t]
	\centering
	\includegraphics[width=\linewidth]{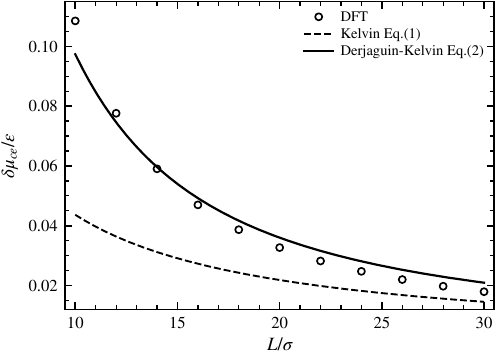}
	\caption{\label{fig:CE_inf_slit}
		Capillary evaporation shift $\delta\mu_{ce}$ in an infinite slit formed by two hard walls at $T=0.92T_c$. Symbols denote DFT results. The dashed line is the Kelvin prediction, Eq.~(\ref{kelvin}), and the solid line is the Derjaguin-Kelvin equation, Eq.~(\ref{kelvin_derj}), with $n=2$ appropriate for SR forces. The Derjaguin correction gives a nearly quantitative description of the DFT data over most of the slit widths considered.}
\end{figure} 

Minimization of $\Omega[\rho]$ gives the Euler-Lagrange equation
\begin{equation}
	k_BT\ln[\rho({\bf r})\Lambda^3]+\frac{\delta {\cal F}_{\rm ex}}{\delta\rho({\bf r})}+V({\bf r})-\mu=0,
	\label{eq:DFT_EL}
\end{equation}
where $V({\bf r})$ is either $V_{\rm LR}(x,z)$ or $V_{\rm SR}(x,z)$, and which is solved iteratively on a two-dimensional grid in the $x$-$z$ plane. Capillary condensation in the LR system and capillary evaporation in the SR system are located by comparing the grand-potential branches obtained from two different initial density profiles, corresponding to the gas-like and liquid-like states, with the transition identified by the crossing of these branches.

\begin{figure*} 
	\centering 
	\begin{tabular}{ccc} 
		\includegraphics[width=0.48\textwidth]{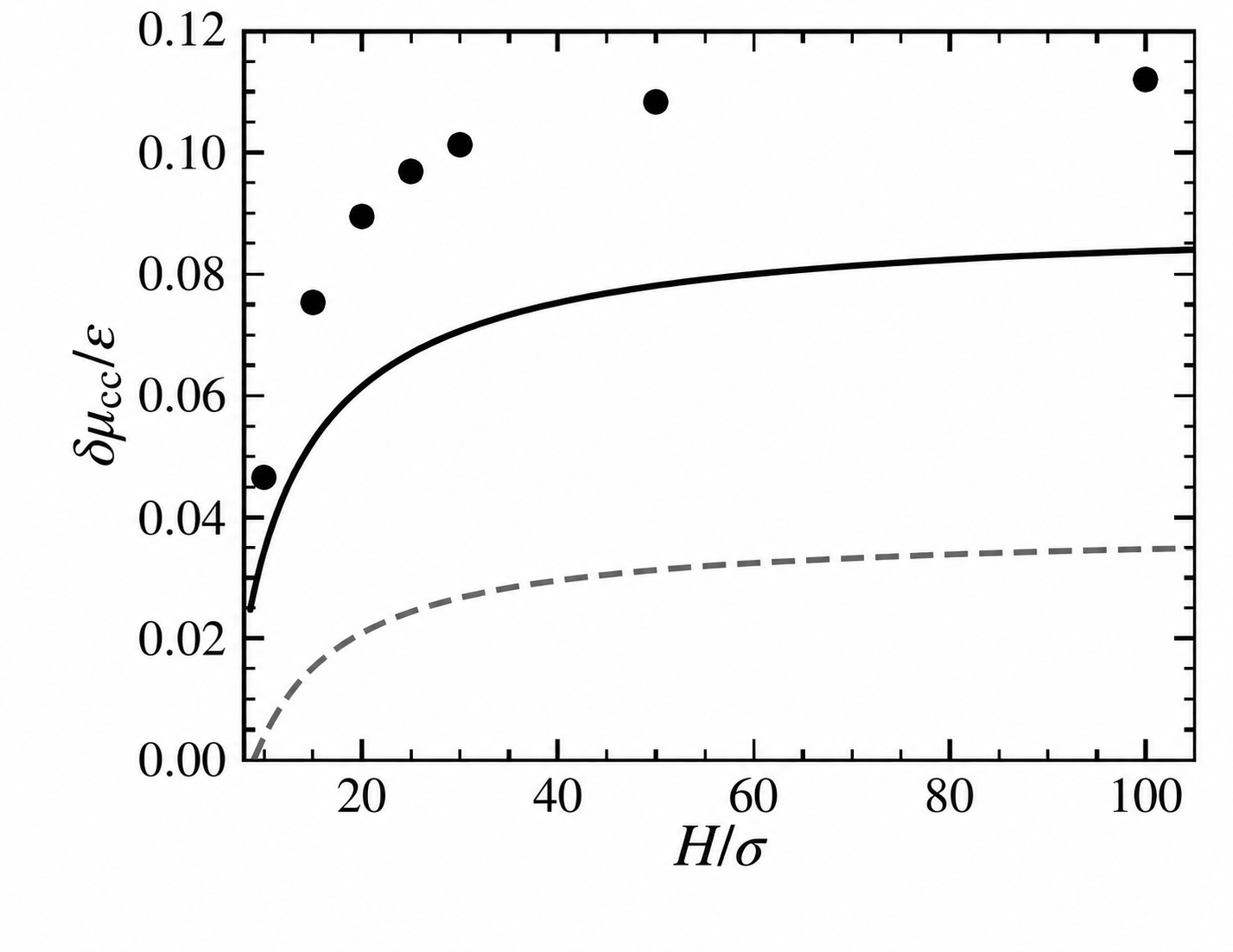} & \includegraphics[width=0.49\textwidth]{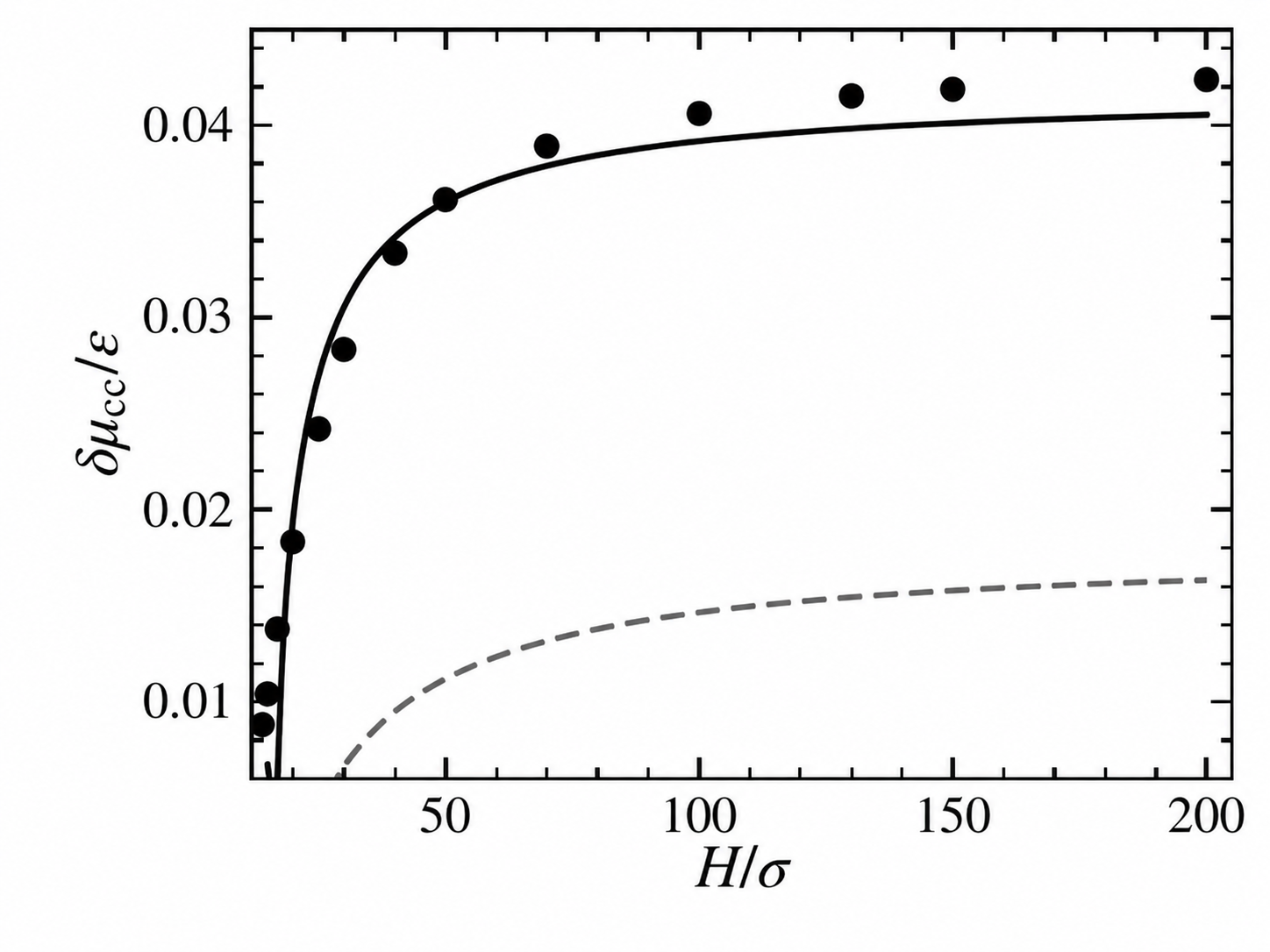} \\ 
		[-1mm] (a) $L=10\sigma$ & 
		(b) $L=20\sigma$ \\
	\end{tabular}
	\caption{\label{fig:LR_finite_slits}
		Capillary condensation shift $\delta\mu_{cc}$ in finite slits with attractive Lennard-Jones walls at $T=0.92T_c$, shown as a function of the wall length $H$ for two slit widths: $L=10\sigma$ and $20\sigma$. Symbols denote DFT results, dashed curves correspond to the finite-slit Kelvin equation without wetting-layer corrections, and solid curves show the Derjaguin-corrected prediction, Eq.~\eqref{eq:DGK_scaling_general}.
	}
\end{figure*} 

All numerical results presented below are obtained at the fixed temperature $T=0.92T_c$, where the critical temperature of the fluid model is $k_BT_c/\varepsilon=1.41$. For the attractive-wall model, for which we set   $\rho_w\varepsilon_w=\varepsilon\sigma^{-3}$, this temperature is above the wetting temperature $T_w\simeq0.8T_c$, so that the walls are completely wet and $\theta=0$. At the same time, the temperature is sufficiently far below $T_c$ for bulk critical effects to be negligible.  At this temperature the coexisting bulk densities are $\rho_g=0.104\sigma^{-3}$ and $\rho_l=0.431\sigma^{-3}$. The corresponding liquid-gas surface tension, obtained independently from a planar free interface between the coexisting bulk gas and liquid phases, is $\gamma=0.0561\,\varepsilon\sigma^{-2}$.

 \begin{figure*}[t]
	\centering
	\begin{tabular}{cc} 
		\includegraphics[width=0.48\textwidth]{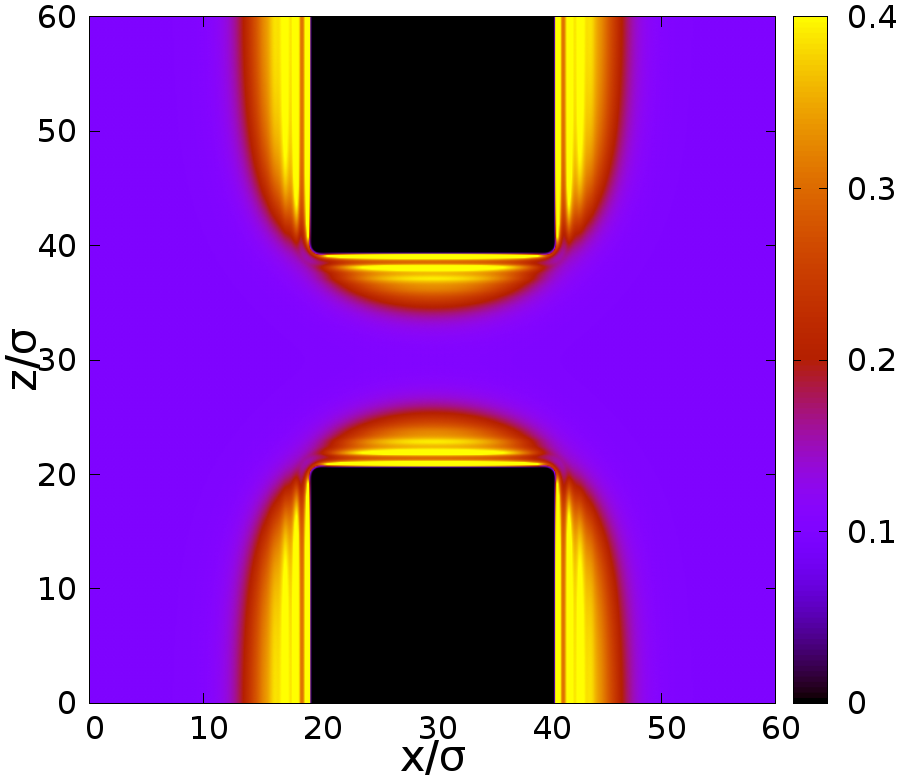} & \includegraphics[width=0.48\textwidth]{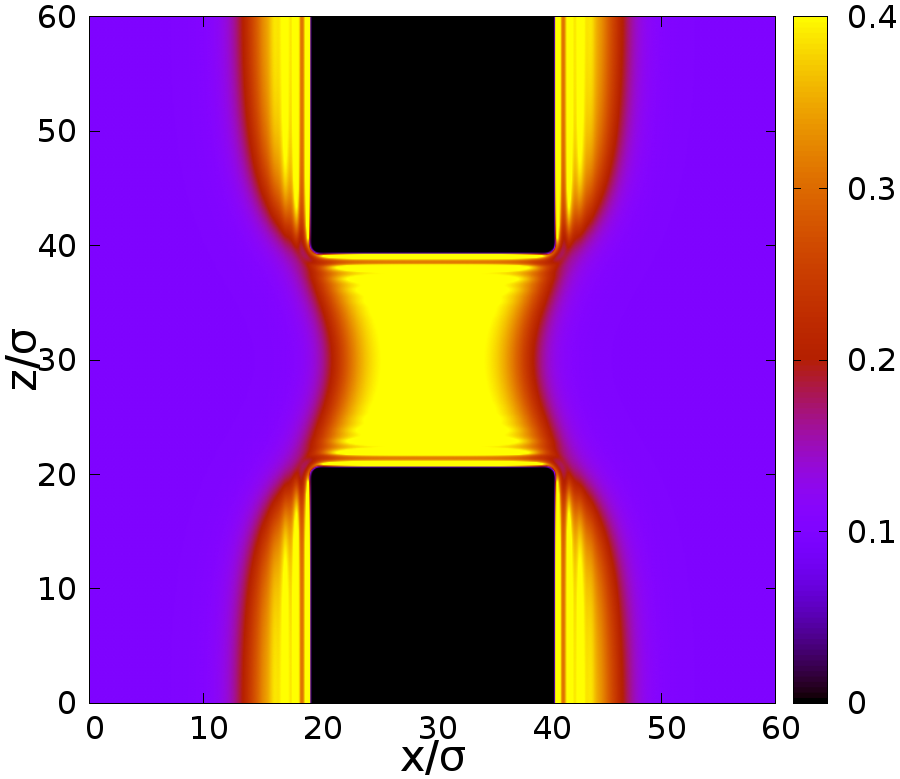} \\\\
		[-1mm] (a) $H=20\sigma$, low-density state & 
		(b) $H=20\sigma$, high-density state \\
		[2mm] \includegraphics[width=0.48\textwidth]{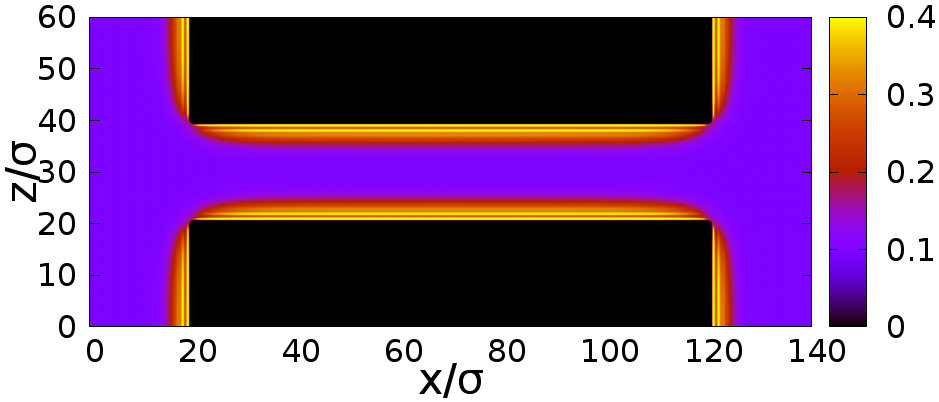} & \includegraphics[width=0.48\textwidth]{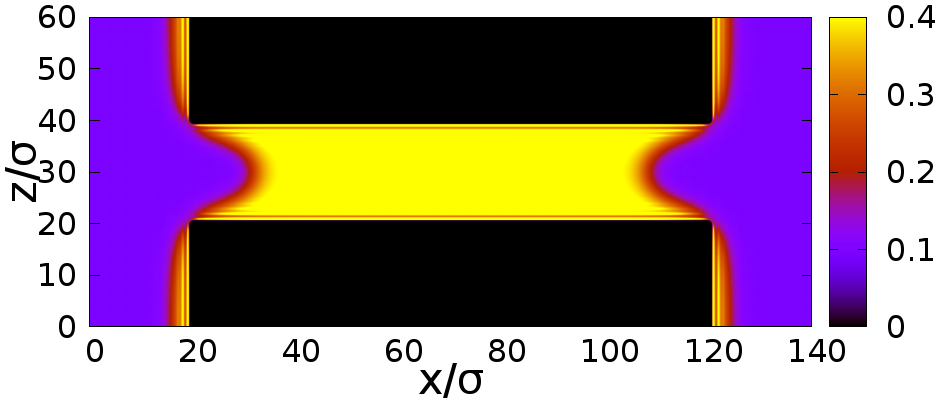} \\\\
		[-1mm] (c) $H=100\sigma$, low-density state & 
		(d) $H=100\sigma$, high-density state \end{tabular}
	\caption{\label{fig:LR_profiles}
		Coexisting DFT density profiles for the LR system with $L=20\sigma$. The upper panels correspond to $H=20\sigma$, while the lower panels correspond to $H=100\sigma$. For each wall length, the low-density state shows the adsorbed liquid structure at the walls, while the high-density state shows the capillary liquid state. For $H=20\sigma$ the adsorbed film is droplet-like, since the two edge regions overlap. For $H=100\sigma$ the adsorbed liquid forms extended film-like layers along the walls, corresponding to the film regime.}
\end{figure*} 

\begin{figure*} [t]
	\centering 
	\begin{tabular}{ccc} 
		\includegraphics[width=0.48\textwidth]{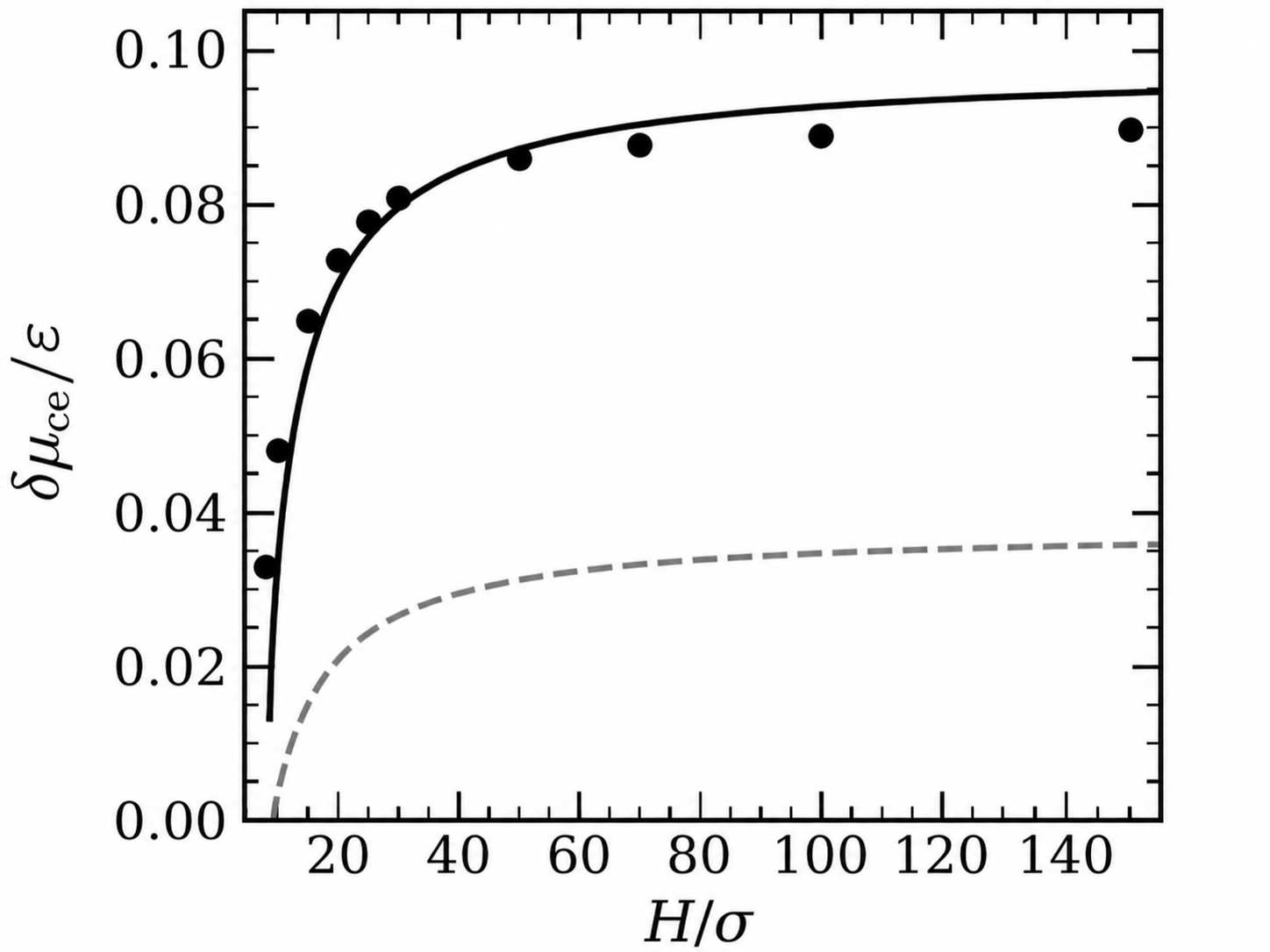} & \includegraphics[width=0.48\textwidth]{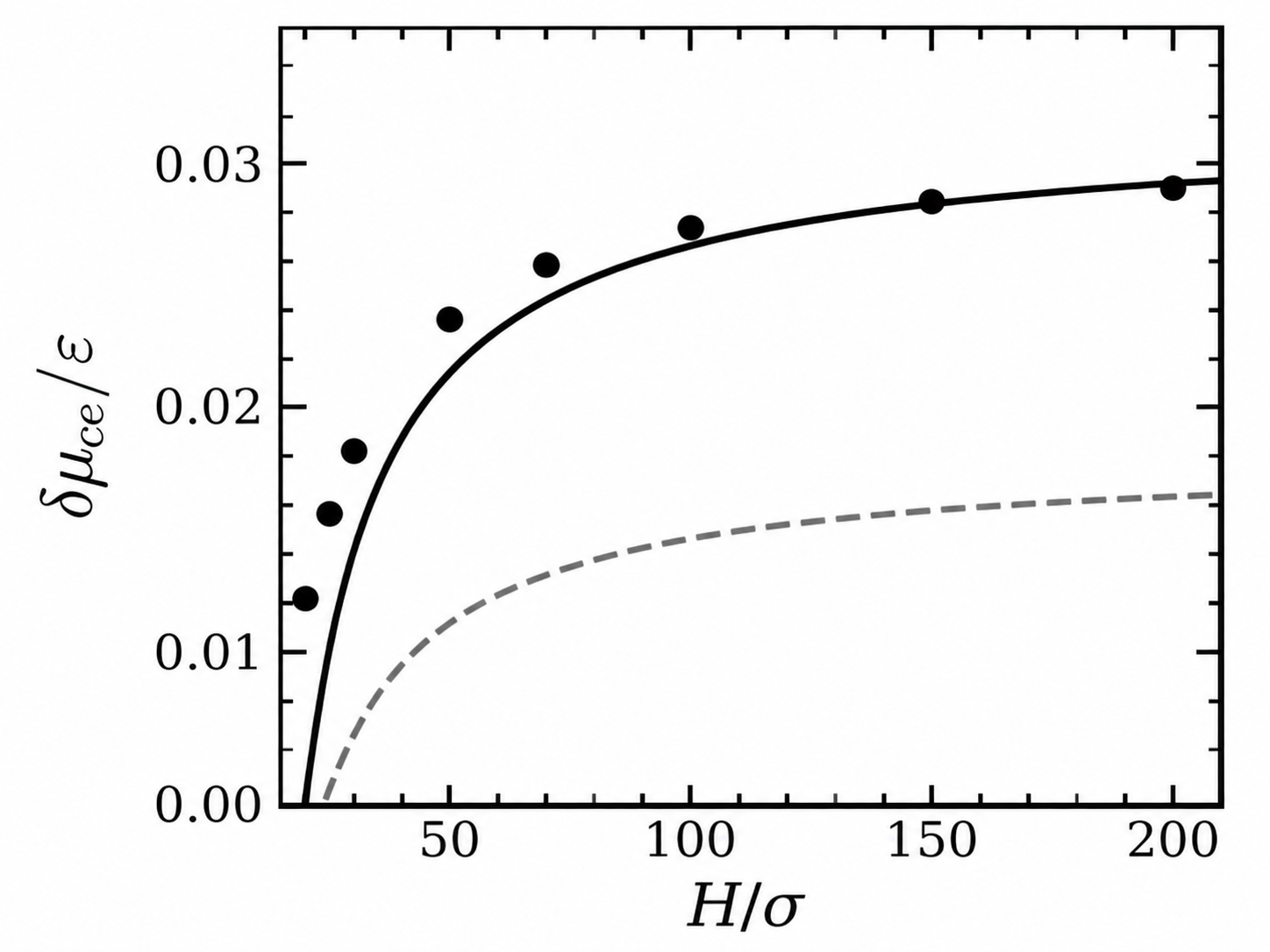} \\ 
		[-1mm] (a) $L=10\sigma$ & 
		(b) $L=20\sigma$ \\
	\end{tabular}
	\caption{\label{fig:SR_finite_slits}
		Capillary evaporation shift $\delta\mu_{ce}$ in finite slits with hard walls at $T=0.92T_c$, shown as a function of the wall length $H$ for two slit widths: $L=10\sigma$ and $20\sigma$. Symbols denote DFT results, dashed curves correspond to the finite-slit Kelvin equation without wetting-layer corrections, and solid curves show the Derjaguin-corrected prediction, Eq.~\eqref{eq:DGK_scaling_general}. 
	}
\end{figure*} 

\begin{figure*}[t]
	\centering
	\begin{tabular}{cc} 
		\includegraphics[width=0.48\textwidth]{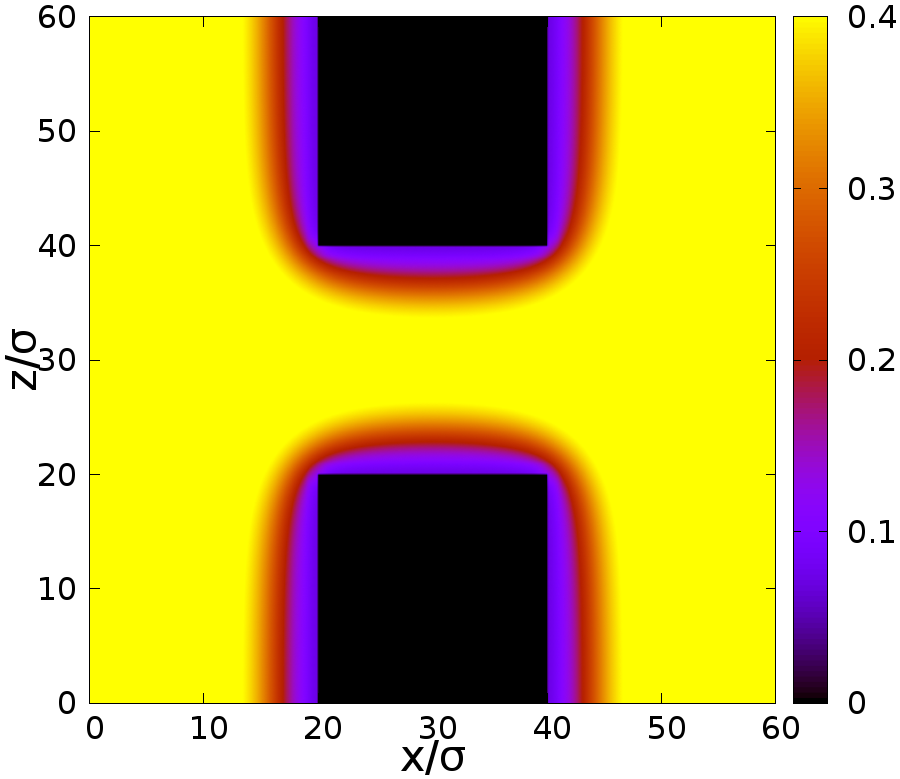} & \includegraphics[width=0.48\textwidth]{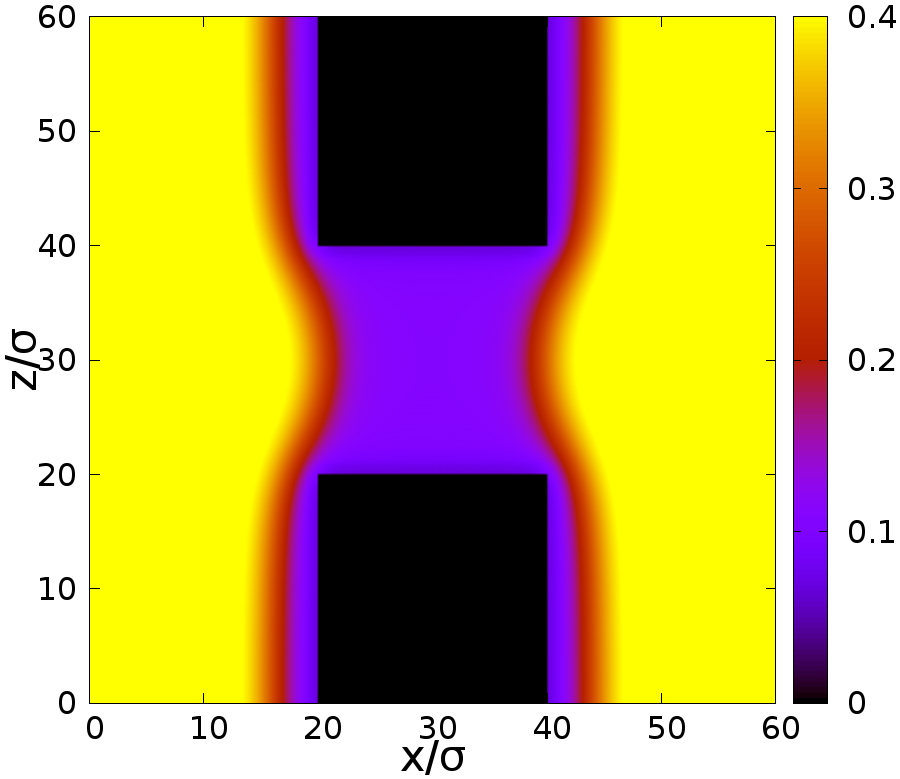} \\\\
		[-1mm] (a) $H=20\sigma$, high-density state & 
		(b) $H=20\sigma$, low-density state \\
		[2mm] \includegraphics[width=0.48\textwidth]{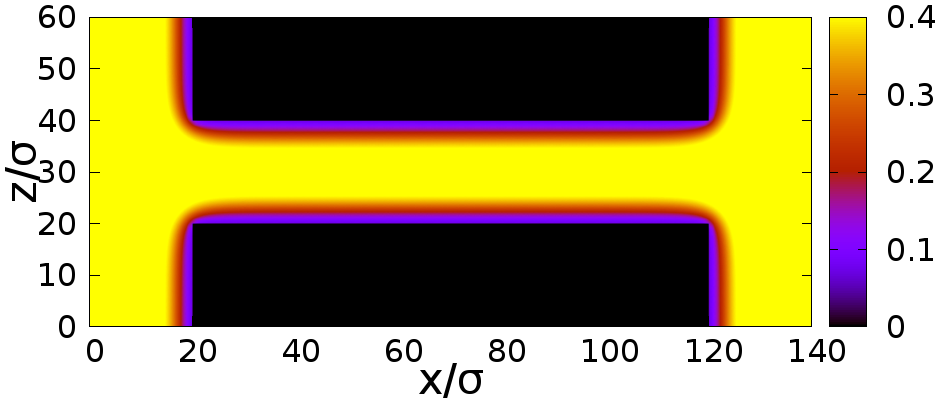} & \includegraphics[width=0.48\textwidth]{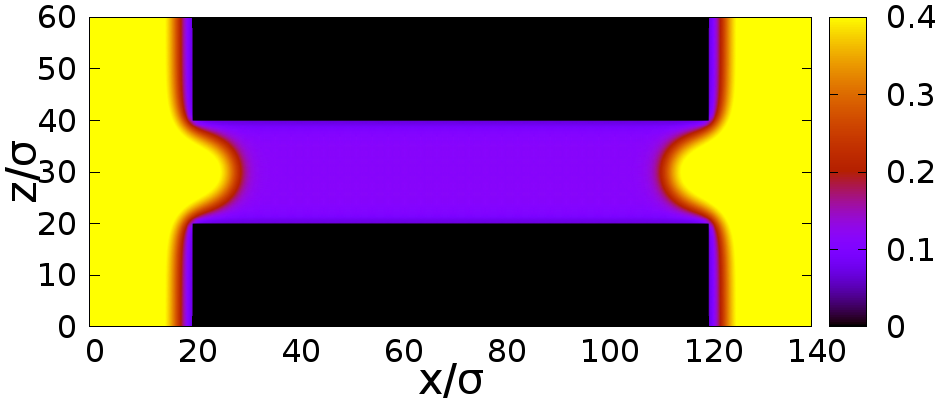} \\\\
		[-1mm] (c) $H=100\sigma$, high-density state & 
		(d) $H=100\sigma$, low-density state \end{tabular}
	\caption{\label{fig:SR_profiles}
		Coexisting DFT density profiles for the SR system with $L=20\sigma$. The upper panels correspond to $H=20\sigma$, while the lower panels correspond to $H=100\sigma$. For each wall length, the high-density state shows the adsorbed gas structure at the walls, while the low-density state shows the capillary-evaporated  state. For $H=20\sigma$ the adsorbed gas film is droplet-like, since the two edge regions overlap. For $H=100\sigma$ the adsorbed gas forms extended film-like layers along the walls, corresponding to the film regime.}
\end{figure*}

 \subsection{Capillary condensation and evaporation in finite slits}
 
 We start by presenting planar reference calculations for simple wall-fluid systems, which are needed to determine the input function $\ell_\pi(\delta\mu)$ entering the generalized Derjaguin-Kelvin equation. Specifically, we consider two one-dimensional problems at $T=0.92T_c$: complete wetting at an attractive Lennard-Jones $9$-$3$ wall, approached as $\mu\to\mu_{\rm sat}^{-}$, and complete drying at a hard wall, approached as $\mu\to\mu_{\rm sat}^{+}$. In both cases the density profile varies only in the direction perpendicular to the wall, and the resulting planar film thicknesses are shown in Fig.~\ref{fig:ell_pi}.

 For the Lennard-Jones wall, the log-log plot confirms the expected asymptotic complete-wetting law $\ell_\pi\sim\delta\mu^{-\beta_s^{\rm co}}$, with $\beta_s^{\rm co}=1/3$ for LR dispersion forces. Farther from saturation, the DFT data deviate from the leading asymptotic form. However, including the first subdominant corrections,
 \begin{equation}
 	\ell_\pi(\delta\mu)=a_0\delta\mu^{-1/3}+a_1+a_2\delta\mu^{1/3},
 	\label{eq:ellpi_LR_fit}
 \end{equation}
in units of $\sigma$ and $\varepsilon$ with $a_0=1.163$, $a_1=3.051$ and $a_2=-10.408$, provides an accurate fit over the whole range of chemical potentials considered; see Fig.~\ref{fig:ell_pi}(a). Comparing the leading term of Eq.~\eqref{eq:ellpi_LR_fit} with the asymptotic relation \eqref{eq:ell_LR_single} and using $\delta p\simeq\Delta\rho\delta\mu$, with $\Delta\rho=\rho_l-\rho_g=0.327\sigma^{-3}$, gives $A/\varepsilon=a_0^3\Delta\rho\sigma^3/2\simeq0.257$.
 
For the hard-wall system, the drying layer is expected to grow logarithmically, $\ell_\pi(\delta\mu)\simeq-\xi_b\ln\delta\mu$, where $\xi_b$ is the bulk correlation length of the gas phase at the temperature considered. As shown in Fig.~\ref{fig:ell_pi}(b), the semi-log plot confirms that this asymptotic dependence with $\xi_b=1.184\sigma$ describes the DFT data very accurately over the entire interval of $\delta\mu$.

As a next step, we consider infinitely long slits and determine how the capillary condensation or evaporation shift depends on the slit width. These one-dimensional calculations provide the $H\to\infty$ reference limit for the finite slits considered below and serve as a consistency check of the numerical procedure.

Figure~\ref{fig:CC_inf_slit} shows the capillary condensation shift in an infinite slit formed by two planar Lennard-Jones walls. The DFT results are compared with the classical Kelvin equation, Eq.~(\ref{kelvin}), and with the Derjaguin-Kelvin equation, Eq.~(\ref{kelvin_derj}), using $n=3$ appropriate for LR dispersion forces. The comparison shows that the Kelvin equation substantially underestimates the condensation shift, especially for narrow slits. Including the Derjaguin correction considerably improves the agreement, reflecting the importance of the wetting films adsorbed at the walls. Some discrepancy remains at the smallest slit widths, where microscopic packing effects and non-asymptotic corrections to the mesoscopic description are expected to be most pronounced.\\

The analogous calculation for the hard-wall slit is shown in Fig.~\ref{fig:CE_inf_slit}. In this case the walls are completely dry and the relevant transition is capillary evaporation from the liquid side. The DFT values of the evaporation shift are compared with the Kelvin equation and with the Derjaguin-Kelvin equation with $n=2$, as appropriate for SR forces. Again, the Kelvin equation underestimates the magnitude of the shift, whereas the Derjaguin correction brings the theoretical prediction much closer to the DFT results. For the hard-wall system the agreement is nearly perfect over a broad range of slit widths.

We now turn to the main numerical test of the finite-slit theory. We focus first on the LR system, where the slit is formed by two finite Lennard-Jones walls and the relevant transition is capillary condensation from the gas side. DFT calculations were performed for two slit widths, $L=10\sigma$ and $L=20\sigma$, and for several wall lengths $H$. For each pair $(L,H)$, the condensation chemical potential was determined from the crossing of the gas-like and liquid-like grand-potential branches.

The results are shown in Fig.~\ref{fig:LR_finite_slits}, where the DFT data are compared with the finite-slit Kelvin equation and with its Derjaguin-corrected form, Eq.~\eqref{eq:DGK_scaling_general}. 
The comparison shows that the Derjaguin correction substantially improves the agreement with DFT for both slit widths. The remaining deviations are largest for the narrowest slit, $L=10\sigma$, where microscopic packing effects and non-asymptotic corrections to the mesoscopic description are expected to be most important.

To illustrate the microscopic origin of the two limiting regimes entering the theory, Fig.~\ref{fig:LR_profiles} shows representative coexisting density profiles for $L=20\sigma$. For the shorter wall, $H=20\sigma$, the low-density branch contains a droplet-like adsorbed liquid structure: the wetting layers nucleated at the two open edges overlap before a planar film can be established. This corresponds to the regime $H\lesssim\xi_\parallel$, where the finite-wall residual free energy is cut off by $H$. By contrast, for $H=100\sigma$ the adsorbed liquid in the low-density state forms extended film-like layers along the walls, with a nearly planar central region. This is the film regime, $H\gg\xi_\parallel$, in which the two edge regions are effectively independent. The coexisting high-density profiles show the corresponding capillary liquid states separated from the surrounding gas by terminal menisci pinned at the slit openings.

We finally test the SR version of the finite-slit Derjaguin-Kelvin equation. The SR system is represented by slits formed by purely hard walls, embedded in a bulk liquid. The relevant transition is therefore capillary evaporation rather than capillary condensation, in which case the roles  of the gas-like and liquid-like phases interchange.

The results are shown in Fig.~\ref{fig:SR_finite_slits}  for two slit widths, $L=10\sigma$ and $L=20\sigma$. The DFT data are compared with the finite-slit Kelvin equation, which contains only the macroscopic meniscus contribution, and with the Derjaguin-corrected prediction for SR forces. As in the LR case, the Derjaguin correction substantially improves the agreement with DFT. Here the agreement is particularly good even for the narrower slit, $L=10\sigma$, reflecting the fact that the Derjaguin-Kelvin equation already provides a nearly quantitative description of the corresponding infinite hard-wall slit. The improvement is again not limited to the $H\to\infty$ limit; the corrected theory also follows the observed finite-$H$ dependence of the evaporation shift much more closely than the purely macroscopic Kelvin equation.

Representative coexisting density profiles for the SR system are shown in Fig.~\ref{fig:SR_profiles} for $L=20\sigma$. The high-density branch corresponds to the liquid-like state before capillary evaporation; in this state the hard walls are coated by gas-like drying layers. The low-density branch corresponds to the capillary-evaporated state. For the shorter wall, $H=20\sigma$, the gas-like drying layers form a droplet-like structure, with strongly overlapping edge regions. For $H=100\sigma$, the gas adsorbed at the hard walls forms extended film-like layers with an almost planar central region, corresponding to the film regime. These profiles provide the SR analogue of the droplet-to-film crossover illustrated above for the LR system.

\section{Concluding remarks}

In this work we have formulated a Derjaguin-corrected Kelvin equation for capillary condensation in finite open slits. The starting point was the macroscopic generalized Kelvin equation, in which the effect of the finite wall length is reflected by the presence of symmetric menisci pinned at the slit ends. Although this equation determines the condensation shift only implicitly through the edge contact angle, we showed that it can be  represented with high accuracy  by a simple scaling approximation. This approximation is constrained by the exact limiting behaviours at both ends of the interval of aspect ratios and provides a closed analytic expression for the finite-slit condensation shift.

The main result of the paper is the mesoscopic extension of this construction to completely wet walls. In this regime, the confining walls in the low-density gas-like state contain adsorbed wetting layers whose free-energy contribution affects the phase boundary. By analyzing the wetting layer on a single finite wall, we decomposed the residual finite-size contribution into an edge free energy, $2\tau_e$, and a Casimir-like interaction, ${\cal{F}}_C$, between the two edge regions. This decomposition makes explicit the competition between the wall length $H$ and the parallel correlation length $\xi_\parallel$: for $H\gg\xi_\parallel$ the two edges are independent, whereas for $H\lesssim\xi_\parallel$ their relaxation regions overlap. For LR dispersion forces this leads to a logarithmic residual contribution cut off by the smaller of $H$ and $\xi_\parallel$, while for SR forces the residual finite-size term remains microscopic and the dominant correction is the planar Derjaguin contribution.

The resulting generalized Derjaguin-Kelvin equation has the correct limiting structure. In the limit $H\to\infty$ it reduces to the standard Derjaguin-Kelvin equation for an infinite slit, with the effective-width correction $L-n\ell_\pi$, where $n=3$ for LR forces and $n=2$ for SR forces. Conversely, when the adsorbed layers are absent, it reduces to the macroscopic finite-slit Kelvin equation. Combining the mesoscopic equation with the scaling approximation gives a particularly transparent form, Eq.~\eqref{eq:DGK_scaling_general}, and its explicit LR and SR versions, Eqs.~\eqref{eq:DGK_scaling_LR} and \eqref{eq:DGK_scaling_SR}. These equations retain the essential finite-length and wetting-layer contributions while avoiding the need to solve for the edge contact angle explicitly.

The theoretical predictions were tested against microscopic density functional theory calculations. For attractive Lennard-Jones walls, representing the LR complete-wetting regime, the Derjaguin correction substantially improves the prediction of capillary condensation in finite slits compared with the purely macroscopic Kelvin theory. This improvement is largely due to the more accurate $H\to\infty$ limit, but the corrected equation also captures the observed dependence on the wall length. The remaining discrepancies are most visible for the narrowest slits, where microscopic packing effects and subleading corrections to the mesoscopic theory are expected to be strongest.

For the SR systems, represented here by  slits formed of hard walls exhibiting capillary evaporation, the agreement between the Derjaguin-corrected theory and DFT is particularly accurate, even for narrow slits. This reflects the fact that the standard Derjaguin-Kelvin equation already gives an excellent description of the infinite hard-wall slit, and that the finite-length scaling correction captures the dominant additional effect of the terminal menisci. The density profiles obtained from DFT further support the expected picture underlying the theory, showing a crossover from droplet-like adsorbed structures for short walls to film-like adsorbed layers for long walls.

The  present theory is not limited to planar finite slits. Previous studies of bridging transitions between non-planar walls have shown that the bridged region can be mapped onto an effective open slit, whose length, width, and edge contact angle are determined by the local wall geometry \cite{Malijevsky2024}. This construction provides a Kelvin-like description of bridging transitions in simple non-planar geometries and has also been applied to sinusoidally shaped pores \cite{Malijevsky2024a}. Since the effective finite slit is the central element of this mapping, the Derjaguin-corrected equation developed here provides a natural mesoscopic extension of such approaches. In particular, it offers a way to include wetting- or drying-layer effects in the complete wetting and complete drying regimes, where the purely macroscopic mapping is no longer sufficient. The present theory may therefore serve as the basis for a more accurate description of bridging transitions in curved, corrugated, or patterned confinements.

Another natural extension concerns slits with competing or
opposing walls, for which the two surfaces preferentially adsorb
different bulk phases. This situation is qualitatively different
from the like walls considered here, since the competing surface preferences can require a spatially varying bulk field, rather than a uniform shift of $\mu$, in order to maintain two-phase coexistence in a finite system. In particular, as shown by Carlon
and Drzewinski~\cite{Carlon1997} for the two-dimensional
Ising strip with opposing surface fields, a spatially varying
bulk field, interpretable for fluids as a gravitational field,
can restore two-phase coexistence up to the bulk critical point
and leads to a distinct finite-size scaling problem. Extending
the present Derjaguin-corrected finite-slit construction to such
competing-wall geometries would require incorporating this
field gradient together with the wetting- or drying-layer
corrections, and is left for future work.

%

\end{document}